\documentclass[12pt]{article}
\usepackage{hyperref}
\usepackage[top=2cm,bottom=3cm,left=2cm,right=2cm]{geometry}
\usepackage{graphicx}
\usepackage{amsmath}
\usepackage{amssymb}

\numberwithin{equation}{section}
\numberwithin{figure}{section}

\def\eq#1{(\ref{eq:#1})}

\def\d{\partial}

\def\UHP{\mathrm{UHP}}
\def\BCFT{\mathrm{BCFT}}

\begin{document}

\begin{titlepage}

	\hfill \today
	\begin{center}
		\vskip 2cm
		
		{\Large \bf Symplectic structure in open string field theory I: \\ Rolling tachyons}

		\vskip 0.5cm
		
		\vskip 1.0cm
		{\large {Vin\'{\i}cius Bernardes$^{1}$, Theodore Erler$^{1}$, and Atakan Hilmi F{\i}rat$^{2}$ }}
		
		\vskip 0.5cm
		
		{\em  \hskip -.1truecm
			$^{1}$
			CEICO, FZU - Institute of Physics of the Czech Academy of Sciences \\
			No Slovance 2, 182 21, Prague 8, Czech Republic
			\\
			\vskip 0.5cm
			$^{2}$
			Center for Quantum Mathematics and Physics (QMAP),
			Department of Physics \& Astronomy, \\
			University of California, Davis, CA 95616, USA
			\\
			\vskip 0.5cm
			\tt \href{mailto:viniciusbernsilva@gmail.com}{viniciusbernsilva@gmail.com},
		\href{mailto:tchovi@gmail.com}{tchovi@gmail.com},
		 \href{mailto:ahfirat@ucdavis.edu}{ahfirat@ucdavis.edu} \vskip 5pt }
		
		\vskip 2.0cm
		{\bf Abstract}
		
	\end{center}
	\vskip 0.25cm
	\noindent
	\begin{narrower}
		\baselineskip15pt
		
		\noindent We discuss a new formula for the symplectic structure on the phase space of open string field theory. Revisiting the setup of Cho, Mazel, and Yin, we use the formula to compute the energy of rolling tachyon solutions on unstable D-branes. An important aspect of the analysis is dealing with the singular ultraviolet behavior of string vertices in Lorentzian signature, a feature we refer to as {\it transgressive locality}. This forces us to carry out computations in momentum space, where time and causality are somewhat obscure. Nevertheless the symplectic structure appears to be sensible, giving results in agreement with boundary state computations. As further confirmation of our methods, we study the symplectic structure for rolling tachyons in scalar effective field theory, where vertices show similar high energy behavior to string field theory but the physics is that of local field theory. This model gives interesting insight into   the runaway oscillations of the rolling tachyon.

	\end{narrower}
\end{titlepage}

\tableofcontents
\baselineskip15pt

\section{Introduction}

In recent work \cite{CovL} we proposed a formula for the symplectic structure on the phase space of an arbitrary Lagrangian field theory:
\begin{equation}
\Omega = \frac{1}{2}\omega\big(\delta\Phi,[Q_\Phi,\sigma]\delta\Phi\big).\label{eq:Omega}
\end{equation}
The formula assumes that the theory has been expressed in terms of a cyclic $L_\infty$ algebra, where $\omega$ is the Batalin-Vilkovisky inner product, $Q_\Phi$ is the kinetic operator around a solution $\Phi$, and $\sigma$ is a grade zero operator called the {\it sigmoid}. This is the first of three papers \cite{Bernardes1,Bernardes2} whose goal is to test this formula in the context of string field theory (SFT). The hope is that the symplectic structure can serve as a foundation for a Hamiltonian description of the theory, which may give access to new observables in string theory such as energy, topological charge, and black hole entropy. In principle the formula \eq{Omega} applies to both open and closed string field theory. However, we will focus on open (bosonic) strings. Closed SFT is a theory of gravity, which makes extraction of observables (such as energy) more subtle. Some relevant work on diffeomorphism invariance and boundary terms in closed SFT can be found in \cite{Erler7,Stettinger2,Firat2,Maccaferri2,Maccaferri3,Mamade1,Mamade2,Hull,Mazel,Frenkel}. We leave applications to closed SFT for the future. 

An important motivation for our work is the analysis of Cho, Mazel, and Yin \cite{Cho} who use an older proposal for the symplectic structure due to Witten \cite{Witten} to compute the energy of rolling tachyon solutions on unstable D-branes \cite{Sen2}. Their results match with boundary state calculations carried out by Kudrna \cite{Kudrna}. The agreement, however, requires discarding certain infinities contained within Witten's formula. In this paper we revisit this calculation using the new symplectic structure. Our ability to confirm these results is a minimum standard for the formalism we are trying to develop. 

The divergences in Witten's symplectic structure stem from the use of the open string midpoint to measure time. In our calculations, we will measure time with the open string center of mass. This is the time coordinate associated to the component fields of the Fock space expansion of the SFT action. With this definition of time, the symplectic form \eq{Omega} is finite. A further benefit is that its evaluation does not rely on special properties of the Witten vertex that have no analogue in other SFTs. A challenge, however, is that Witten's string field theory is highly nonlocal in the center of mass coordinate. The nature of the nonlocality is also strange. The interactions are in a sense ``more than local'' in time, as expressed by couplings which grow faster than exponentially at high energies. We refer to this as {\it transgressive locality.} A consequence is that fields do not interact in a well-defined way as a function of time. Instead interactions must be expressed in momentum space. The lack of a clear picture of time evolution is worrisome given that we are trying to do Hamiltonian mechanics, but we will see that everything is working. 

As confirmation of our methods we test the symplectic structure in a simpler model. We consider an effective description of scalar $\phi^3$ theory derived by attaching {\it stubs} to the interactions. Normally the effective field theory would be defined in Euclidean signature, but here we are interested in Lorentzian signature where the model exhibits transgressive locality in a very similar way as SFT. Scalar $\phi^3$ theory has an unstable vacuum with rolling tachyon solutions of various energies. We derive the counterpart of these solutions in the effective field theory, and use the symplectic structure \eq{Omega} to determine their energies to the first subleading order. Despite the unusual nature of the theory's nonlocality, the energies come out correctly. The effective field theory also gives an interesting point of view on the significance of rolling tachyon oscillations seen in $p$-adic string theory and open SFT~\cite{Moeller,Hata,Erbin2}.  

Finally we consider open string field theory. We compute the symplectic structure for rolling tachyon solutions in Siegel gauge using the center of mass coordinate to measure time. We reduce the expression to first subleading order to a collection of correlation functions on the surface defining the off-shell 4-point amplitude in Siegel gauge. The correlation functions are evaluated using squeezed state oscillator methods developed by Taylor \cite{Taylor}, showing good agreement with the results of Cho, Mazel and Yin and very close agreement with boundary state results of Kudrna. The details of our numerical implementation can be found in a {\tt Mathematica} file accompanying this paper. We conclude by reviewing how the divergences of Witten's symplectic structure can be eliminated by using the lightcone coordinate of the string midpoint to measure time \cite{Maeno,Erler2}. Independence from the choice of sigmoid can then be used to explain why our results agree with those of Cho, Mazel, and Yin.

\subsubsection*{Conventions}

We use mostly plus metric and $\alpha'=1$. Fourier transform conventions are
\begin{equation}
f(x) = \int \frac{d^Dk}{(2\pi)^D} e^{ik\cdot x}f(k),\ \ \ \ \ f(k) = \int d^D x \, e^{-ik\cdot x}f(x),
\end{equation}
where $f(x)$ is a function in position space and $f(k)$ is its Fourier transform in momentum space. The ghost correlator is normalized as 
\begin{equation}\langle c(z_1) c(z_2) c(z_3)\rangle_\UHP^\mathrm{gh} = z_{12}z_{13}z_{23},\ \ \ z_{ij}=z_i-z_j,
\end{equation}
and we use the left handed convention for the open string star product \cite{Erler3}. Commutators are always graded with respect to Grassmann parity.

\section{Transgressive locality}

Witten's SFT can be truncated to the tachyon field $T$, where the action takes the form
\begin{align}
S = \int d^D x \left(\frac{1}{2}T(\Box+1)T - \frac{\kappa}{3}\left(e^{\frac{\Lambda}{2}\Box}T\right)^3\right),\label{eq:Taction}
\end{align}
where $D=26$, $\Box= \d_\mu \d^\mu$ and
\begin{align}
\kappa = \left(\frac{27}{16}\right)^{3/2},\ \ \ \ \Lambda = \ln\left(\frac{27}{16}\right)\approx 0.52.\label{eq:kappaLambda}
\end{align}
The cubic term contains an infinite derivative operator $e^{\frac{\Lambda}{2}\Box}$ which appears (for some $\Lambda$) ubiquitously in interactions between any fields in the Fock space expansion of any covariant string field theory.  For this reason nonlocality in string field theory takes a fairly universal form. The infinite derivative operator can be interpreted as a formal power series in $\Box$.  However this definition is strictly perturbative in $\alpha'$. A more complete definition can be given in terms of a position space integral transform,
\begin{align}
e^{\frac{\Lambda}{2}\Box}T(x)  = (2\pi \Lambda)^{-D/2} \int d^D x' \exp\left(-\frac{(x-x')^2}{2\Lambda}\right) T(x'),\label{eq:pos}
\end{align}
or a momentum space Fourier transform 
\begin{align}
e^{\frac{\Lambda}{2}\Box}T(x) = \int \frac{d^D k}{(2\pi)^D} e^{i k\cdot x} e^{-\frac{\Lambda}{2} k^2} T(k),\label{eq:mom}
\end{align}
where $T(k)$ is the Fourier transform of $T(x)$. In Euclidean signature these definitions are equivalent. However, since we are trying to develop the Hamiltonian formalism, it seems necessary to use Lorentzian signature. In this context the two definitions are not equivalent and both seem problematic. The position space integral transform has an inverted Gaussian in the time direction. This creates divergence unless the field vanishes very quickly at large times. The Fourier transform has an inverted Gaussian at high energies. This creates divergence unless high energy modes are very strongly suppressed.  

In the covariant phase space formalism we work on-shell, and solutions are not expected to vanish at large times. Therefore the position space representation cannot work. The momentum space representation might work if solutions vanish fast enough at high energy. We do not know whether they do, but for example a plane wave solution is a delta function in energy, which certainly vanishes fast enough. The issue of high energies in SFT has been discussed by Pius and Sen~\cite{Pius} in the context of loop amplitudes. The inverted Gaussian appears to create severe ultraviolet divergence in integration over loop momenta. The resolution is to pin the endpoints of the energy contour to~$\pm i\infty$ while analytically continuing external momenta. The prescription suggests that Lorentzian signature should be understood through some kind of analytic continuation from Euclidean signature. However, we will not proceed in this way. Instead we employ the momentum space representation and work in Lorentzian signature directly. 

The problem with momentum space is that we cannot talk directly about time evolution. This could be worrisome because the consistency of the symplectic structure relies on the assumption that fields propagate and interact in a somewhat localized fashion as a function of time. It might appear that SFT interactions are not like this. However, we will argue that really they are. In fact, in a certain sense, fields interact only if they are {\it closer} than coincident in time. This sounds contradictory, but it demonstrates the difficulty with the position space picture and why it is not expected to be an issue for us. The covariant phase space formalism is primarily concerned with infrared physics, and the nature of interactions at short distances should not matter. The singular nature of SFT interactions in time will be called {\it transgressive locality.} 

To understand what transgressive locality means it is useful to look at the interaction of wavepackets. The cubic coupling of tachyons in Witten's SFT defines a notion of multiplication 
\begin{equation}
f_1\star f_2 = e^{\frac{\Lambda}{2}\Box}\Big((e^{\frac{\Lambda}{2}\Box}f_1)(e^{\frac{\Lambda}{2}\Box}f_2)\Big).\label{eq:f1sf2}
\end{equation}
We consider wavepackets which are localized functions of time only. The goal is to measure how rapidly the product becomes small as the wavepackets are separated. The overall size of the product can be quantified as 
\begin{equation}
||f_1\star f_2||^2 = \int_{-\infty}^\infty dt\, |f_1\star f_2|^2.\label{eq:g12}
\end{equation}
To compute this, the wavepackets must be chosen so that their Fourier transforms vanish fast enough at high energy. This forbids, for example, wavepackets with compact support. We consider Gaussians: 
\begin{equation}f_1(t) = \frac{1}{\sqrt{2\pi}\sigma}e^{-\frac{(t-t_1)^2}{2\sigma^2}},\ \ \ \ \ \ f_2(t) = \frac{1}{\sqrt{2\pi}\sigma} e^{-\frac{(t-t_2)^2}{2\sigma^2}}.\end{equation}
The first Gaussian is centered at $t=t_1$ and the second at $t=t_2$. The product \eq{f1sf2} converges if the Gaussians are sufficiently spread out in time,
\begin{equation}\sigma^2>3\Lambda.\end{equation}
Using \eq{mom} we can confirm that 
\begin{equation}
e^{\frac{\Lambda}{2}\Box}f_1 = \frac{1}{\sqrt{2\pi}\sqrt{\sigma^2-\Lambda}} \exp\left[-\frac{(t-t_1)^2}{2(\sigma^2-\Lambda)}\right].
\end{equation}
The width of the Gaussian is narrowed. As the Gaussians are narrowed, one expects that their product will decrease more rapidly with separation. Computing \eq{g12} leads to 
\begin{equation}||f_1\star f_2|| = \left(\frac{1}{32\pi^3(\sigma^2-\Lambda)}\right)^{1/4}\exp\left[-\frac{(t_1-t_2)^2}{4(\sigma^2-\Lambda)}\right].\end{equation}
The interaction decreases as a Gaussian function of the separation. The rate of decrease is {\it faster} than it is with local, pointwise multiplication of wavepackets (which would correspond to $\Lambda=0$). This is the sense that SFT interactions are ``more local" in time than in a local theory. In spatial directions, however, effectively $\Lambda<0$ and the interaction decreases more slowly with separation than with pointwise multiplication. This is closer to the traditional concept of nonlocality as representing ``action at a distance."

\section{Rolling tachyon in effective scalar field theory}

We write the $\phi^3$ action as 
\begin{equation}
\widehat{S} = \int d^D x \left(-\frac{1}{2}\d_\mu\widehat{\phi}  \d^\mu\widehat{\phi} + \frac{\mu^2}{2}\widehat{\phi}^2 -\frac{g}{3}\widehat{\phi}^3\right),
\end{equation}
where $\widehat{\phi}$ is the (bare) scalar field, $-\mu^2$ is the tachyon mass-squared and $g$ is the cubic coupling constant. We study an effective description of this theory where particles travel a worldline distance $\Lambda/2$ inside the cubic vertex before joining at an interaction point. The propagation of each particle inside the vertex is called a {\it stub}, and $\Lambda/2$ is called the {\it stub length}. Stubs have been discussed extensively in string field theory \cite{Sen,Brustein,Sachs2,Schnabl,Schnabl2,Stettinger,Erbin,Maccaferri,Firat}, but also in ordinary field theories in \cite{Costello,Sachs,Erler}. The stubs define a cutoff scale for the effective field theory. We write the effective scalar field as $\phi$, and the effective action expanded to quartic order is \cite{Erler}
\begin{align}
S & = \int d^D x \left(-\frac{1}{2}\d_\mu\phi \d^\mu\phi + \frac{\mu^2}{2}\phi^2 -\frac{g}{3}\big(e^{\frac{\Lambda}{2}(\Box +\mu^2)}\phi\big)^3\right.\nonumber\\
&  \ \ \ \ \ \ \ \ \ \ \ \ \ \ \ \ \left.+\frac{g^2}{2}\big(e^{\frac{\Lambda}{2}(\Box+\mu^2)}\phi\big)^2 \int_0^\Lambda ds\, e^{s(\Box+\mu^2)} \big(e^{\frac{\Lambda}{2}(\Box+\mu^2)}\phi\big)^2 + \, \mathcal{O}(g^3) \right).
\end{align}
The effective action has a quartic vertex which is necessary to account for 4-point processes where the intermediate particle travels a worldline distance shorter than $\Lambda$. Each field in the interaction comes with a factor $e^{\frac{\Lambda}{2}(\Box+\mu^2)}$ which we call the {\it stub operator}. It represents Euclidean evolution on the worldline over a distance~$\Lambda/2$. The stub operator also appears in the Fock space expansion of the string field theory action, as can be seen in \eq{Taction}. From this it is clear that the effective field theory exhibits exactly the same kind of nonlocality as seen in string field theory. 

We are interested in time dependent, rolling tachyon solutions in the effective field theory. The question is whether the energy of these solutions as determined by the symplectic structure \eq{Omega} agrees with what can be inferred by mapping the solutions from the original $\phi^3$ theory. This serves as a nontrivial check on the consistency of our methods. 
 
\subsection{Rolling scalar field and its energy}

The hyperbolic cosine rolling tachyon solution of $\phi^3$ theory can be written as a perturbative expansion in a parameter~$\widehat{\lambda}$ 
\begin{equation}
\widehat{\phi}(t) = \widehat{\lambda} \widehat{\phi}_1(t) + \widehat{\lambda}^2 \widehat{\phi}_2(t) + \widehat{\lambda}^3\widehat{\phi}_3(t) +\, \mathcal{O}(\widehat{\lambda}^4).
\end{equation}
Up to order $\widehat{\lambda}^3$ the equations of motion give the expressions
\begin{subequations}
\begin{align}
\widehat{\phi}_1(t) & = \cosh(\mu t),\\[.2cm]
\widehat{\phi}_2(t) & = \frac{g}{\mu^2}\left(\frac{1}{2}-\frac{1}{6}\cosh(2\mu t)\right),\\
\widehat{\phi}_3(t) & = \frac{g^2}{\mu^4}\left(\frac{1}{48}\cosh(3\mu t)-\frac{5}{12}\mu t \sinh(\mu t)\right).
\end{align}
\end{subequations}
This represents a time-dependent solution where the scalar rolls up the potential and stops just short of the unstable vacuum $\hat{\phi}=0$ before turning around.  We shift the origin so that the turnaround occurs at $t=0$. The deformation parameter $\widehat{\lambda}$ is defined so as to ensure that the first hyperbolic cosine harmonic appears only at first order in~$\widehat{\lambda}$. At order $\widehat{\lambda}^3$ there is a contribution from the hyperbolic sine (times $t$).  This happens because---in terminology borrowed from SFT---the deformation is not exactly marginal.  The third order equation of motion,
\begin{equation}(\Box+\mu^2)\widehat{\phi}_3 = 2g\widehat{\phi}_1\widehat{\phi}_2,\end{equation}
has a term on the right hand side which is in the kernel of the Klein-Gordon operator $(\Box+\mu^2)$. This term can be referred to as an ``obstruction" to exact marginality. In open SFT, the rolling tachyon solution is exactly marginal \cite{Callan,Polchinski} and can be written using hyperbolic cosine harmonics alone. 

Solutions of $\phi^3$ theory map to solutions of the effective field theory through the stub operator~\cite{Schnabl,Erler}
\begin{equation}
\phi = e^{\Lambda(\Box+\mu^2)} \widehat{\phi}.\label{eq:phiphihat}
\end{equation}
In this way we can find the rolling tachyon solution in the effective field theory  
\begin{equation}
\phi(t) = \lambda \phi_1(t) + \lambda ^2 \phi_2(t) + \lambda^3 \phi_3(t)+\, \mathcal{O}(\lambda^4),\label{eq:eff_rolling1}
\end{equation}
where 
\begin{subequations}\label{eq:eff_rolling2}
\begin{align}
\phi_1(t) & = \cosh(\mu t),\\[.2cm]
\phi_2(t) & = \frac{g}{\mu^2}\left(\frac{e^{-\frac{\Lambda\mu^2}{2}}}{2}-\frac{e^{-\frac{3\Lambda\mu^2}{2}}}{6}\cosh(2\mu t)\right),\\
\phi_3(y) & = \frac{g^2}{\mu^4}\left(\frac{e^{-4\Lambda\mu^2}}{48}\cosh(3\mu t)-\frac{5}{12}\mu t \sinh(\mu t)\right).
\end{align}
\end{subequations}
The parameter $\lambda$ is defined so that the first hyperbolic cosine harmonic appears only at first order in $\lambda$. This implies a nontrivial relation to the parameter $\widehat{\lambda}$ of the original solution
\begin{equation}
\lambda = \widehat{\lambda} -\frac{5}{12}\frac{\Lambda g^2}{\mu^2} \widehat{\lambda}^3 +\, \mathcal{O}(\widehat{\lambda}^5).\label{eq:lambda_Lambda}
\end{equation}
The energy of the rolling tachyon in $\phi^3$ theory can be determined by the depth of the potential at the turnaround point, where the kinetic energy vanishes. The value of the scalar field at the turnaround point ($t=0$) is 
\begin{equation}\widehat{\phi}(0) = \widehat{\lambda} + \frac{1}{3}\frac{g}{\mu^2}\widehat{\lambda}^2 + \frac{1}{48}\frac{g^2}{\mu^4}\widehat{\lambda}^3+\mathcal{O}(\widehat{\lambda}^4).\end{equation}
and substituting into the potential gives
\begin{equation}
V\big(\widehat{\phi}(0)\big) = -\frac{\mu^2}{2}\widehat{\phi}(0)^2+\frac{g}{3}\widehat{\phi}(0)^3 = -\frac{\mu^2}{2}\widehat{\lambda}^2+\frac{37}{144}\frac{g^2}{\mu^2}\widehat{\lambda}^4+\, \mathcal{O}(\widehat{\lambda}^6).
\end{equation}
Writing this as a function of $\lambda$ and multiplying by the volume $V$ of the spatial dimensions, we can infer the energy of the rolling tachyon solution in the effective field theory:
\begin{equation}
E(\lambda) = V \left[-\frac{\mu^2}{2}\lambda^2+\frac{1}{4}\left(\frac{37}{36}+\frac{5}{3}\Lambda\mu^2\right)\frac{g^2}{\mu^2}\lambda^4+\, \mathcal{O}(\lambda^6)\right].\label{eq:eff_E}
\end{equation}
Our objective is to reconstruct this result (up to order $\lambda^4$) by evaluating the symplectic form \eq{Omega} directly in the effective field theory using the solution \eq{eff_rolling1}-\eq{eff_rolling2}.

\subsection{Symplectic structure}

The symplectic structure of the effective field theory can be derived following the procedure described in \cite{CovL}. The starting point is the free action for a fluctuation $\varphi$ around a background solution~$\phi$,
\begin{equation}
S_\mathrm{fluctuation} = -\frac{1}{2}\int d^D x\, \varphi\big(Q_\phi\varphi\big),
\end{equation}
where $Q_\phi$ is the appropriate (nonlocal) wave operator depending on $\phi$. Up to second order $g$ the wave operator is 
\begin{align}
Q_\phi & = -\big(\Box+\mu^2\big) \nonumber\\
&\ \ \  +2g\, e^{\frac{\Lambda}{2}(\Box+\mu^2)}\big(e^{\frac{\Lambda}{2}(\Box+\mu^2)}\phi\big)e^{\frac{\Lambda}{2}(\Box+\mu^2)} \nonumber\\
&\ \ \  -4 g^2\, e^{\frac{\Lambda}{2}(\Box+\mu^2)}\big(e^{\frac{\Lambda}{2}(\Box+\mu^2)}\phi\big)\int_0^\Lambda ds\, e^{s(\Box+\mu^2)} \big(e^{\frac{\Lambda}{2}(\Box+\mu^2)}\phi\big)e^{\frac{\Lambda}{2}(\Box+\mu^2)} \nonumber\\
& \ \ \ - 2g^2\, e^{\frac{\Lambda}{2}(\Box+\mu^2)} \left( \int_0^\Lambda ds\, e^{s(\Box+\mu^2)}\big(e^{\frac{\Lambda}{2}(\Box+\mu^2)}\phi\big)^2\right) e^{\frac{\Lambda}{2}(\Box+\mu^2)}\nonumber\\
&\ \ \ +\, \mathcal{O}(g^3),
\end{align}
where the stub operators act on everything that follows to the right until parentheses close. The symplectic structure of the effective field theory is given by
\begin{equation}\Omega = -\frac{1}{2}\int d^D x\, \delta\phi [Q_\phi,\sigma] \delta\phi, \end{equation}
where the operator $\sigma$ is called the {\it sigmoid} and satisfies boundary conditions
\begin{equation}\lim_{t\to-\infty}\sigma = 0,\ \ \ \ \lim_{t\to\infty}\sigma=1.\label{eq:boundary_conditions}\end{equation}
We will assume that the sigmoid acts through multiplication by a function $\sigma(x)$ on spacetime. 
It is important that the commutator $[Q_\phi,\sigma]$ is evaluated before it is inserted into the symplectic form. Otherwise the formula is ambiguous because a divergence from integration over time multiplies a vanishing factor from the constraint 
\begin{equation}Q_\phi\delta\phi = 0,\end{equation}
which follows because $\delta$ preserves the equations of motion. The ambiguity can also be understood as an issue of specifying the correct boundary terms in the symplectic form \cite{CovL}. Unfortunately, we cannot meaningfully evaluate the commutator $[Q_\phi,\sigma]$ in position space because of transgressive locality. We will therefore write the symplectic form in momentum space. Up to second order in $g$ we find
\begin{align}
\Omega & = \frac{1}{2}\int \frac{d^Dk^0 d^Dk^1 d^Dk^2}{(2\pi)^{3D}}(2\pi)^D\delta^D(k^{012})\Big[\sigma(k^0)\Big((k^{01})^2-(k^1)^2\Big)\Big]\delta\phi(k^1)\delta\phi(k^2)\nonumber\\
& \ \ \ +g\int \frac{d^Dk^0 d^Dk^1 d^Dk^2 d^Dk^3}{(2\pi)^{4D}}(2\pi)^D\delta^D(k^{0123})\nonumber\\
&\ \ \ \ \ \ \ \ \ \ \ \ \times\left[\sigma(k^0)\left(e^{-\frac{\Lambda}{2}((k^{012})^2+(k^{01})^2+(k^2)^2-3\mu^2)}-\left(\scriptsize{\begin{matrix}\text{same with} \\ k^0=0\end{matrix}}\right)\right)\right]\delta\phi(k^1)\phi(k^2)\delta\phi(k^3)\nonumber\\
& \ \ \ -g^2\int \frac{d^Dk^0 d^Dk^1 d^Dk^2 d^Dk^3 d^D k^4}{(2\pi)^{5D}}(2\pi)^D\delta^D(k^{01234})\nonumber\\
&\ \ \ \ \ \ \ \ \ \ \ \ \times\Bigg[\sigma(k^0)\Bigg(e^{-\frac{\Lambda}{2}((k^{0123})^2+(k^{01})^2+(k^2)^2+(k^3)^2-4\mu^2)}\left(2\frac{1-e^{-\Lambda((k^{013})^2-\mu^2)}}{(k^{013})^2-\mu^2}+\frac{1-e^{-\Lambda((k^{23})^2-\mu^2)}}{(k^{23})^2-\mu^2}\right)\nonumber\\
&\ \ \ \ \ \ \ \ \ \ \ \ \ \ \ \ \ \ \ \ \ \ \ \ \ \ \ \ \   -\left(\scriptsize{\begin{matrix}\text{same with} \\ k^0=0\end{matrix}}\right)\Bigg)\Bigg]\delta\phi(k^1)\phi(k^2)\phi(k^3)\delta\phi(k^4)\nonumber\\
& \ \ \ +\, \mathcal{O}(g^3).\label{eq:Omega_eff_k}
\end{align}
Here $\sigma(k),\delta\phi(k),\phi(k)$ are the Fourier modes of $\sigma(x),\delta\phi(x),\phi(x)$ and we combine momentum labels to abbreviate
\begin{equation}k^{ABC\cdots} = k^A+ k^B + k^C +\cdots.\end{equation}
The boundary conditions \eq{boundary_conditions} constrain the form of the sigmoid towards zero momentum 
\begin{equation}\sigma(k) \sim (2\pi)^{D-1}\delta^{D-1}(\vec{k})\times \frac{1}{i E} \ \ \text{as}\ \ k_\mu \to 0,\label{eq:k_boundary_conditions}\end{equation}
where $E=k_0$ is the timelike component of the momentum (the energy) and $\vec{k}$ are the spatial components. The delta function in spatial momenta reflects the fact that the sigmoid is homogeneous along spatial dimensions in the infinite  past and future. The pole at zero energy ensures that the sigmoid differs by $1$ between the infinite past and future. Since the symplectic structure is supposed to be independent of the sigmoid, and elsewhere the sigmoid is arbitrary, it follows that all contributions to the symplectic structure must originate from zero momentum. The pole at zero energy creates ambiguity in multiplication of distributions. A prototypical example is the product 
\begin{equation}  \frac{1}{iE} \, i E \, \delta(E),\end{equation}
where $\delta(E)$ might appear after substituting an explicit solution into the symplectic form. Multiplying $i E$ to the left gives $\delta(E)$ while multiplying it to the right gives zero. This is the ambiguity mentioned earlier whose resolution is to evaluate the commutator $[Q_\phi,\sigma]$ before inserting it into the symplectic form.  In momentum space, this means that the sigmoid must be multiplied inside the square brackets in~\eq{Omega_eff_k},
\begin{equation}\Big[\sigma(k^0)\Big( \cdots\Big)\Big],\end{equation}
before anything else. It can be seen in \eq{Omega_eff_k} that $\sigma(k^0)$ multiplies a function of the momenta which vanishes at $k^0=0$. This cancels the pole of the sigmoid, eliminating any ambiguity in computation of the symplectic form. 

The rolling tachyon is a perturbative, homogeneous, time-dependent solution. Such a solution is part of a two dimensional phase space differing by the size of the deformation parameter $\lambda$ and the origin of the time coordinate~$t_0$. The differential in this phase space takes the form 
\begin{align}
\delta \phi(t) & = \delta\lambda \Big[\phi_1(t) + 2\lambda\phi_2(t) + 3 \lambda^2 \phi_3(t)+\, \mathcal{O}(\lambda^3)\,\Big]\nonumber\\
&\ \ \ -\delta t_0 \Big[\lambda \dot{\phi}_1(t) + \lambda^2 \dot{\phi}_2(t) + \lambda^3 \dot{\phi}_3(t)+\mathcal{O}(\lambda^4)\Big],
\end{align}
where dot denotes the derivative with respect to time. We used
\begin{equation}
\frac{\d}{\d t_0}\phi(t) = -\dot{\phi}(t),
\end{equation}
since shifting the origin of time forwards is the same as shifting time backwards. The symplectic structure admits a perturbative expansion
\begin{equation}
\Omega = \delta t_0 \delta\lambda\Big[\lambda \Omega_1+\lambda^2 \Omega_2 +\lambda^3\Omega_3 +\mathcal{O}(\lambda^4)\Big].
\end{equation}
Let us assume that the sigmoid is a function of time only $\sigma(t)$. We further integrate out spatial coordinates to produce a factor of the spatial volume $V$. Finally, we Fourier transform so that integration over time is replaced with integration over energy. In this way, the leading coefficient of the symplectic form is found to be
\begin{align}
\Omega_1 = V \int \frac{dE^0dE^1dE^2}{(2\pi)^3} 2\pi\delta(E^{012}) \dot{\sigma}(E^0)K_2(E^0,E^1,E^2)\dot{\phi}_1(E^1)\phi_1(E^2),\label{eq:eff_Omega1}
\end{align}
where $\sigma(E)$ and $\phi_n(E)$ are Fourier modes of $\sigma(t)$ and $\phi_n(t)$ and we introduce 
\begin{equation}
K_2(E^0,E^1,E^2) = \frac{(E^{01})^2-(E^1)^2}{iE^0} =-i(E^1-E^2).\label{eq:K2}
\end{equation}
The arguments of $K_2$ are assumed to add to zero by momentum conservation. We use the dot to denote the time derivative, which in momentum space amounts to multiplication by $iE$. In the integrand we have multiplied and divided by $iE^0$ so that the sigmoid is replaced by its time derivative. The boundary condition \eq{k_boundary_conditions} implies that the time derivative satisfies 
\begin{equation}\lim_{E\to 0}\dot{\sigma}(E)=1.\label{eq:sigma_om_bc}\end{equation}
Concerning the prefactor $K_2(E^0,E^1,E^2)$ it is important that the pole at $E^0=0$ is canceled against the numerator before multiplication with anything else. Because the numerator is polynomial in the energy, the cancellation can be carried out explicitly as in \eq{K2}. In higher order prefactors the numerators are not polynomial, and the cancellation cannot be carried out explicitly. Instead we must fill in a removable singularity with the value of the limit. The second order coefficient can be written as a sum of terms,
\begin{equation}\Omega_2 = \Omega_2^{(2)}+\Omega_2^{(3)},\label{eq:eff_Omega2}\end{equation}
which are respectively quadratic and cubic in the field:
\begin{subequations}
\begin{align}
\Omega_2^{(2)} &= V \int \frac{dE^0dE^1dE^2}{(2\pi)^3} 2\pi\delta(E^{012}) \dot{\sigma}(E^0)K_2(E^0,E^1,E^2) \Big(\dot{\phi}_2(E^1)\phi_1(E^2)+ 2\dot{\phi}_1(E^1)\phi_2(E^2)\Big),\label{eq:eff_Omega22}\\
\Omega_2^{(3)}& =-2g V \int \frac{d E^0 dE^1 dE^2 dE^3}{(2\pi)^4} 2\pi\delta(E^{0123})\dot{\sigma}(E^0)K_3(E^0,E^1,E^2,E^3)\dot{\phi}_1(E^1)\phi_1(E^2)\phi_1(E^3),\label{eq:eff_Omega23}
\end{align}
\end{subequations}
where we introduce
\begin{equation}
K_3(E^0,E^1,E^2,E^3) = \frac{e^{\frac{\Lambda}{2}((E^{012})^2+(E^{01})^2+(E^2)^2+3\mu^2)}-\left(\scriptsize{\begin{matrix}\text{same with} \\ E^0=0\end{matrix}}\right)}{i E^0}.\label{eq:K3}
\end{equation}
The arguments of $K_3$ are assumed to add to zero by momentum conservation. The pole at $E^0=0$ must be canceled against the numerator before multiplication with anything else.
The third order coefficient is a sum of three terms,
\begin{equation}\Omega_3 = \Omega_3^{(2)}+\Omega_3^{(3)}+\Omega_3^{(4)},\label{eq:eff_Omega3}\end{equation}
which are respectively quadratic, cubic, and quartic in the field:
\begin{subequations}
\begin{align}
\Omega_3^{(2)} & = V \int \frac{dE^0dE^1dE^2}{(2\pi)^3} 2\pi\delta(E^{012}) \dot{\sigma}(E^0)K_2(E^0,E^1,E^2)\Big(\dot{\phi}_3(E^1)\phi_1(E^2) + 2\dot{\phi}_2(E^1)\phi_2(E^2)\nonumber\\
&\ \ \ \ \ \ \ \ \ \ \ \ \ \ \ \ \ \ \ \ \ \ \ \ \ \ \ \ \ \ \ \ \ \ \ \ \ \ \ \ \ \ \ \ \ \ \ \ \ \ \ \ \ \ \ \ \ \ \ \ \ \ \ \ \ \ \ \ \ \ 
 +3\dot{\phi}_1(E^1)\phi_3(E^2)\Big),\label{eq:eff_Omega32}\\
\Omega_3^{(3)} & = -2 g V \int \frac{d E^0 dE^1 dE^2 dE^3}{(2\pi)^4} 2\pi\delta(E^{0123})\dot{\sigma}(E^0)K_3(E^0,E^1,E^2,E^3)\nonumber\\
& \ \ \ \ \ \ \ \ \ \ \ \ \ \  
\times\Big(\dot{\phi}_1(E^1)\phi_2(E^2)\phi_1(E^3)+\dot{\phi}_2(E^1)\phi_1(E^2)\phi_1(E^3)+2\dot{\phi}_1(E^1)\phi_1(E^2)\phi_2(E^3)\Big),\label{eq:eff_Omega33}\\
\Omega_3^{(4)} & = -2 g^2 V\int \frac{d E^0 dE^1 dE^2 dE^3 dE^4}{(2\pi)^5} 2\pi\delta(E^{01234})\dot{\sigma}(E^0)K_4(E^0,E^1,E^2,E^3,E^4)\nonumber\\
& \ \ \ \ \ \ \ \ \ \ \ \ \ \  \ \ \ \ \ \ \ \ \ \ \ \ \ \ \ \ \ \ \ \ \ \ \ \ \ \ \ \ \ \ \ \ \ \ \ \ \ \ \ \ \ \ \ \ \ \ \ \ \ \ \ \ \ \ \ \ 
\times \dot{\phi}_1(E^1)\phi_1(E^2)\phi_1(E^3)\phi_1(E^4),\label{eq:eff_Omega34}
\end{align}
\end{subequations}
where we introduce
\begin{align}
& K_4(E^0,E^1,E^2,E^3,E^4) = \nonumber\\
&\ \ \ \ \ \ \ \ \frac{e^{\frac{\Lambda}{2}((E^{0123})^2+(E^{01})^2+(E^2)^2+(E^3)^2+4\mu^2)}\left(2\frac{1-e^{\Lambda((E^{013})^2+\mu^2)}}{(E^{013})^2+\mu^2}+\frac{1-e^{\Lambda((E^{23})^2+\mu^2)}}{(E^{23})^2+\mu^2}\right) -\left(\scriptsize{\begin{matrix}\text{same with} \\ E^0=0\end{matrix}}\right)}{iE^0}.\label{eq:K4}
\end{align}
The arguments of $K_4$ are assumed to add to zero by momentum conservation. The pole at $E^0=0$ must be canceled against the numerator. The energy is determined by writing the symplectic structure as 
\begin{equation}\Omega = \delta t_0 \delta E(\lambda),\end{equation}
from which we infer that 
\begin{equation}
E(\lambda) = \frac{\lambda^2}{2}\Omega_1 + \frac{\lambda^3}{3}\Omega_2 + \frac{\lambda^4}{4}\Omega_3+\mathcal{O}(\lambda^5).
\end{equation}
We evaluate the coefficients $\Omega_1,\Omega_2$ and $\Omega_3$ for the rolling tachyon in the next subsections. 

\begin{figure}[t]
	\centering
	\includegraphics[scale=1.5]{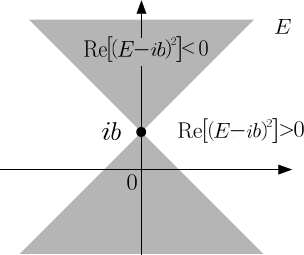}
	\caption{\label{fig:SympOSFT1} The delta function $\delta(E-ib)$ can be realized as the limit \eq{Gaussian} of Gaussian distributions. In the grey region the limit is divergent, while outside the limit vanishes. In this sense the delta function with imaginary argument vanishes for large real energies.}
\end{figure}

The integrands of both $\Omega_2$ and $\Omega_3$ contain factors that grow faster than exponentially at high energies. This is the result of transgressive locality. The symplectic form can be finite and well-defined only if the modes of the perturbative solution vanish fast enough at high energy. The rolling tachyon solution is given in \eq{eff_rolling1}-\eq{eff_rolling2}, which translated to momentum space reads
\begin{subequations}
\begin{align}
\phi_1(E) & = \frac{1}{2}\Big[2\pi \delta(E-i\mu) + 2\pi \delta(E+i\mu)\Big],\\
\phi_2(E) & = \frac{g}{\mu^2}\left[\frac{e^{-\frac{\Lambda\mu^2}{2}}}{2}2\pi\delta(E)-\frac{e^{-\frac{3\Lambda\mu^2}{2}}}{12}\Big(2\pi \delta(E-2i\mu)+2\pi \delta(E+2i\mu)\Big)\right],\\
\phi_3(E) & = \frac{g^2}{\mu^4}\left[\frac{e^{-4\Lambda\mu^2}}{96}\Big(2\pi \delta(E-3i\mu)+2\pi \delta(E+3i\mu)\Big)+\frac{5 i\mu}{24}\Big(2\pi \delta'(E-i\mu)-2\pi \delta'(E+i\mu)\Big)\right].
\end{align}
\end{subequations}
The momentum space solution involves somewhat exotic delta functions with imaginary argument. This is because the hyperbolic sine and cosine are highly non normalizable, and do not have a Fourier transform in the space of ordinary distributions. Presently what matters is whether these delta functions can be argued to vanish for large real energies. To answer this question we define them as a limit of Gaussians:
\begin{equation} 2\pi \delta(E-ib) = \lim_{\sigma\to 0}\frac{\sqrt{2\pi}}{\sigma}\exp\left[-\frac{(E-ib)^2}{2\sigma^2}\right].\label{eq:Gaussian}\end{equation}
Integration should be performed over real energies, but since the Gaussian is analytic everywhere, we can deform the contour without encountering singularity to the line $\mathrm{Im}(E)=b$ where the Gaussian converges to a delta function in the usual sense. The support of the delta function with imaginary argument depends on the sign of the real part of $(E-ib)^2$, as shown in figure \ref{fig:SympOSFT1}. In the quadrants where the sign is negative, the value of the delta function is infinite. In the quadrants where it is positive, the delta function vanishes. Fortunately, large real energies are contained in the region where the delta function vanishes. In this sense the high energy modes of the rolling tachyon solution are very strongly suppressed, and there should be no problems evaluating the symplectic form. 

\subsection{Energy to leading order $\lambda^2$}
\label{subsec:eff_leading}

We calculate the leading coefficient $\Omega_1$. This coefficient is somewhat trivial because it merely determines the energy of a solution to the Klein-Gordon equation. Nevertheless the calculation is useful because it generalizes to higher order coefficients. 

To evaluate $\Omega_1$ we need
\begin{align}
\phi_1(E) & = \frac{1}{2}\Big[2\pi \delta(E-i\mu) + 2\pi \delta(E+i\mu)\Big],\nonumber\\
\dot{\phi}_1(E) & = -\frac{\mu}{2}\Big[2\pi \delta(E-i\mu) - 2\pi \delta(E+i\mu)\Big].
\end{align}
Plugging in to \eq{eff_Omega1} gives
\begin{align}
\Omega_1 & = -\frac{\mu V}{4} \int dE^0dE^1dE^2\delta(E^{012}) \dot{\sigma}(E^0)K_2(E^0,E^1,E^2)\nonumber\\
&\ \ \ \ \ \ \ \ \ \ \ \ \ \ \ \ \ \ \ \ \times\Big(\delta(E^1-i\mu) - \delta(E^1+i\mu)\Big)\Big( \delta(E^2-i\mu) + \delta(E^2+i\mu)\Big).
\end{align}
Multiplying out the delta functions, the terms can be organized into three classes, proportional to the sigmoid evaluated respectively at energies
\begin{equation}-2i\mu,\ \ \ 0,\ \ \ 2i\mu.\end{equation}
Explicitly,
\begin{align}
\Omega_1 = -\frac{\mu V}{4}\!\! \int\! dE^0dE^1dE^2\delta(E^{012})K_2(E^0,E^1,E^2)\bigg[&\dot{\sigma}(-2i\mu)\Big(\delta(E^1-i\mu)\delta(E^2-i\mu)\Big)\nonumber\\
+ ~ &\dot{\sigma}(0)\Big( \delta(E^1\!-\!i\mu)\delta(E^2\!+\!i\mu)\!-\!\delta(E^1\!+\!i\mu)\delta(E^2\!-\!i\mu)\Big)\nonumber\\
+ ~ &\dot{\sigma}(2i\mu)\Big(\delta(E^1+i\mu)\delta(E^2+i\mu)\Big)\bigg].
\end{align}
Evaluating the prefactor for each term we find 
\begin{align}
& K_2(-2i\mu,i\mu,i\mu) = - K_2(2i\mu,-i\mu,-i\mu) = 0, \nonumber\\
& K_2(0,i\mu,-i\mu) = - K_2(0,-i\mu,i\mu)  = 2\mu.
\end{align}
Interestingly, the nonzero energy contributions are multiplied by zero. This is not an accident. The value of the sigmoid at $E = \pm 2i\mu$ can be arbitrary. The only way the symplectic form can be independent from the sigmoid is if these terms come with vanishing coefficient. For the zero energy contributions we use  $\dot{\sigma}(0)=1$ and integrate the delta functions to arrive at
\begin{equation}\Omega_1 = -\mu^2 V.\label{eq:eff_Omega1_result}\end{equation}
This implies that the energy to order $\lambda^2$ agrees with \eq{eff_E}.

From this argument we can learn something about the next coefficient $\Omega_2$. In \eq{eff_Omega22} and \eq{eff_Omega23} we see that this coefficient receives contribution from the product of $\phi_1(E)$ and $\phi_2(E)$ and from the product of three $\phi_1(E)$s. Multiplying out the delta functions the terms organize into four classes, respectively proportional to the sigmoid evaluated at energies
\begin{equation}
3i\mu,\ \ i\mu,\ \ -i\mu,\ \ -3i\mu.
\end{equation}
At all of these energies the value of the sigmoid can be arbitrary. Therefore each class of term must come with vanishing coefficient. Therefore $\Omega_2$ must be zero. We have confirmed this by direct calculation, which involves a cancellation between several terms. 

\subsection{Energy to order $\lambda^4$}

The calculation of $\Omega_3$ is straightforward but lengthy. It is mostly a matter of listing the terms, calculating their coefficients, and adding them up. For the sake of efficiency we discuss only contributions from the sigmoid evaluated at zero energy. We have confirmed that nonzero energy contributions cancel. 

We start with the quadratic contribution \eq{eff_Omega32}. The first step is to extract the zero energy part from the products of $\phi_n$s in the integrand. We find
\begin{align}
    \dot{\phi}_3 (E^1) \phi_1 (E^2) + & 2 \dot{\phi}_2 (E^1)  \phi_2 (E^2) + 3 \dot{\phi}_1 (E^1) \phi_3 (E^2) 
    \nonumber\\[.15cm]
    = \frac{(2\pi)^2 g^2}{\mu^3}  & \bigg[-\frac{5}{48} \Big( \delta(E^1+i\mu) \delta(E^2 - i\mu) - \delta(E^1-i\mu) \delta(E^2 +i \mu)\Big) \nonumber\\
    & -\frac{5i\mu}{48} \Big(\delta' (E^1-i\mu) \delta(E^2 +i \mu) + \delta'(E^1+i\mu) \delta(E^2 - i\mu) \Big) \nonumber\\
    &  + \frac{e^{-3\Lambda\mu^2}}{36} \Big(\delta(E^1+2i\mu) \delta(E^2-2i\mu) - \delta(E^1-2i\mu) \delta(E^2+2i\mu) \Big) \nonumber\\
    &  + \frac{5i\mu}{16} \Big(\delta(E^1-i\mu)  \delta'(E^2+i\mu) + \delta(E^1+i\mu) \delta'(E^2-i\mu) \Big) \bigg] \nonumber\\
    & ~ + \text{nonzero energy}.
\end{align}
Some terms contain a derivative of the delta function, originating from the ``obstruction" to exact marginality in $\phi_3$. After substituting into \eq{eff_Omega22}, the derivatives can be switched to act on the prefactor $K_2(E^0,E^1,E^2)$. We can then integrate the energies and use $\dot{\sigma}(0)=1$ to find
\begin{align}
    \Omega_3^{(2)} =  \frac{g^2 V}{\mu^3} & \bigg[  -\frac{5}{48} \Big(K_2 (0,-i\mu,i\mu) - K_2 (0,i\mu,-i\mu)\Big)\nonumber\\
    &  + \frac{5i\mu}{48}  \frac{d}{ds}\Big(K_2 (0,-i\mu+s,i\mu) + K_2 (0,i\mu+s,-i\mu) \Big)\bigg|_{s=0} \nonumber\\
    &  + \frac{e^{-3\Lambda\mu^2}}{36}(K_2(0,-2i\mu,2i\mu) - K_2 (0,2i\mu,-2i\mu)) \nonumber\\
    &  - \frac{5  i \mu}{16}\frac{d}{ds}\Big( K_2 (0,-i\mu,i\mu+s) + K_2 (0,i\mu,-i\mu+s) \Big)\bigg|_{s=0}  \bigg] + \text{nonzero energy}.
\end{align}
Evaluating the prefactors, 
\begin{subequations}
\begin{align}
&K_2(0,i\mu,-i\mu) = -K_2(0,-i\mu,i\mu) = 2\mu,\\[.2cm] 
& K_2(0,2i\mu,-2i\mu) = -K_2(0,-2i\mu,2i\mu) = 4\mu,\\[.2cm]
&\frac{d}{ds}K_2(0,-i\mu+s,i\mu)\Big|_{s=0} =  \frac{d}{ds}K_2(0,i\mu+s,-i\mu)\Big|_{s=0} = -i,\\
&\frac{d}{ds}K_2(0,-i\mu,i\mu+s)\Big|_{s=0} =  \frac{d}{ds}K_2(0,i\mu,-i\mu+s)\Big|_{s=0} = i,
\end{align}
\end{subequations}
gives
\begin{equation}
\Omega_3^{(2)} = \frac{g^2V}{\mu^2}\left(\frac{5}{4}-\frac{2}{9}e^{-3\Lambda\mu^2}\right)+\text{nonzero\ energy}
\end{equation}
as the result.

The calculation of the cubic and quartic contributions is similar. After extracting the zero energy part of the product of $\phi_n$s and integrating the energy, the cubic contribution is 
\begin{align} 
    \Omega_3^{(3)} = \frac{g^2 V }{\mu} &\bigg[ \frac{e^{\frac{\Lambda\mu^2}{2}}}{2} \Big( K_3 (0,i\mu,-i\mu,0) - K_3 (0,-i\mu,i\mu,0) + \frac{1}{2} K_3 (0,i\mu,0,-i\mu) - \frac{1}{2} K_3 (0,-i\mu,0,i\mu) \Big) \nonumber\\
    & + \frac{e^{-\frac{3\Lambda\mu^2}{2}}}{12} \Big( K_3 (0,-i\mu,-i\mu,2i\mu) - K_3 (0,i\mu,i\mu,-2i\mu) + K_3 (0,-2i\mu,i\mu,i\mu)  \nonumber\\[.1cm]
    & \ \ \ \ \ \ \ \ \ \ \ \  - K_3 (0,2i\mu,-i\mu,-i\mu) + \frac{1}{2} K_3 (-i\mu,2i\mu,-i\mu) - \frac{1}{2} K_3 (i\mu,-2i\mu,i\mu) \Big) \bigg]\nonumber\\
    &  + \text{nonzero energy}.
\end{align}
Evaluating the prefactors,
\begin{subequations}
\begin{align}
& K_3(0,-i\mu,i\mu,0) = -K_3(0,i\mu,-i\mu,0) = -\Lambda\mu e^{\frac{\Lambda\mu^2}{2}},\\
& K_3(0,-i\mu,0,i\mu) = -K_3(0,i\mu,0,-i\mu) = -2\Lambda\mu e^{\frac{\Lambda\mu^2}{2}},\\
& K_3(0,-i\mu,-i\mu,2i\mu) = -K_3(0,i\mu,i\mu,-2i\mu) = -3\Lambda\mu e^{-3\frac{\Lambda\mu^2}{2}},\\
& K_3(0,-2i\mu,i\mu,i\mu) = -K_3(0,2i\mu,-i\mu,-i\mu) = -3\Lambda\mu e^{-3\frac{\Lambda\mu^2}{2}},\\[.2cm]
& K_3(0,-i\mu,2i\mu,-i\mu) = -K_3(0,i\mu,-2i\mu,i\mu) = 0,
\end{align}
\end{subequations}
gives
\begin{align}
\Omega_3^{(3)} = \frac{g^2V}{\mu^2}\Big(2 \Lambda \mu^2 e^{\Lambda\mu^2} - \Lambda\mu^2e^{-3\Lambda\mu^2}\Big)  + \text{nonzero energy}.
\end{align}
The quartic contribution is
\begin{align}
\Omega_3^{(4)}=\frac{g^2\mu V}{8}\bigg[ &- K_4 (0,-i\mu,-i\mu,i\mu,i\mu) + K_4 (0,i\mu,i\mu,-i\mu,-i\mu) \nonumber\\
    &  - K_4 (0,-i\mu,i\mu,-i\mu,i\mu) + K_4 (0,i\mu,-i\mu,i\mu,-i\mu) \nonumber\\
    &  - K_4 (0,-i\mu,i\mu,i\mu,-i\mu) + K_4 (0,i\mu,-i\mu,-i\mu,i\mu) \bigg] + \text{nonzero energy}.
\end{align}
Evaluating the prefactors
\begin{subequations}
\begin{align}
& K_4(0,-i\mu,-i\mu,i\mu,i\mu)=-K_4(0,i\mu,i\mu,-i\mu,-i\mu) = -\frac{6\Lambda}{\mu}\big(1-e^{\Lambda\mu^2}\big),\\
& K_4(0,-i\mu,i\mu,-i\mu,i\mu)=-K_4(0,i\mu,-i\mu,i\mu,-i\mu) \nonumber\\
&\ \ \ \ \ \ \ \ \ \ \ \ \ \ \ \ \ \ \ \ \ \ \ \ \ \ \ \ \ = \frac{1}{\mu^3}\left[\frac{8}{9}-\frac{2}{3}\Lambda\mu^2+ 2\Lambda\mu^2 e^{\Lambda\mu^2}-\left(\frac{8}{9}+4\Lambda\mu^2 \right)e^{-3\Lambda\mu^2} \right],\\
& K_4(0,-i\mu,i\mu,i\mu,-i\mu) = K_4(0,i\mu,-i\mu,-i\mu,i\mu) = 0,
\end{align}
\end{subequations}
gives
\begin{align}
\Omega_3^{(4)} = \frac{V g^2}{\mu^2}\left[-\frac{2}{9}+\frac{5}{3}\Lambda\mu^2 -2\Lambda\mu^2 e^{\Lambda\mu^2}+\left(\frac{2}{9}+\Lambda \mu^2\right)e^{-3\Lambda\mu^2} \right]+\text{nonzero energy}.
\end{align}
In evaluating the prefactors we must take care to correctly cancel the pole in \eq{K3} and \eq{K4}.

Adding everything up,
\begin{align}
\Omega_3  = &~  \Omega_3^{(2)}+\Omega_3^{(3)}+\Omega_3^{(4)}\nonumber\\
 = & ~ \frac{g^2V}{\mu^2}\left[\frac{5}{4}-\frac{2}{9}e^{-3\Lambda\mu^2}\right]\nonumber\\
 & +\frac{g^2V}{\mu^2}\Big[2 \Lambda \mu^2 e^{\Lambda\mu^2} - \Lambda\mu^2e^{-3\Lambda\mu^2}\Big] \nonumber\\
 & + \frac{V g^2}{\mu^2}\left[-\frac{2}{9}+\frac{5}{3}\Lambda\mu^2 -2\Lambda\mu^2 e^{\Lambda\mu^2}+\left(\frac{2}{9}+\Lambda \mu^2\right)e^{-3\Lambda\mu^2} \right]\nonumber\\
= & \frac{g^2V}{\mu^2}\left[\frac{37}{36}+\frac{5}{3}\Lambda\mu^2\right].
\end{align}
Dividing by four gives the order $\lambda^4$ contribution to the energy. Comparing to ~\eq{eff_E} we find agreement.

\subsection{Transgressive locality and runaway oscillations}

At any finite order in perturbation theory, the rolling tachyon solution is made out of a finite combination of (Wick rotated) plane waves. In momentum space, the energy of the plane wave modes is bounded, which means that transgressive locality will not lead to inconsistency in the equations of motion. But beyond perturbation theory, modes of all energies should be present in the solution. Then it is not clear what will happen.

It is not difficult to find nonperturbative solutions in $\phi^3$ theory. An exact solution for the hyperbolic cosine rolling tachyon can be found using elliptic functions. Presently however let us consider the exponential rolling tachyon solution, which is somewhat simpler. It takes the form 
\begin{align}
\widehat{\phi}(t) = \frac{3\mu^2}{2g}\mathrm{sech}^2\left(\frac{\mu(t-t_0)}{2}\right).\label{eq:nonpert_exp}
\end{align}
The solution describes a process where the scalar field falls from the unstable vacuum in the infinite past, bounces off the wall of the cubic potential, and returns to the unstable vacuum in the infinite future. At early times the solution has a perturbative expansion in exponential modes:
\begin{equation}
\widehat{\phi}(t) = \lambda e^{\mu t} -\frac{g}{3\mu^2} \lambda^2 e^{2\mu t} +\frac{g^2}{12\mu^4}\lambda^3 e^{3\mu t} - \mathcal{O}(\lambda^4),\label{eq:pert_exp}
\end{equation}
where $\lambda = 6\mu^2e^{-\mu t_0}/g$. To map this solution to the effective field theory we need to express it in momentum space as $\widehat{\phi}(E)$. Applying the stub operator then gives 
\begin{equation}
\phi(E) = \frac{6\pi}{g}\frac{E}{\sinh\left(\frac{\pi E}{\mu}\right)}e^{\Lambda(E^2+\mu^2)}.\label{eq:nonperturb_eff}
\end{equation}
This does not vanish at high energies. In fact, it grows without bound. Therefore, the rolling tachyon solution does not exist within the space of fields where the action and equations of motion are well-defined. 

\begin{figure}
	\centering
	\includegraphics[scale=.7]{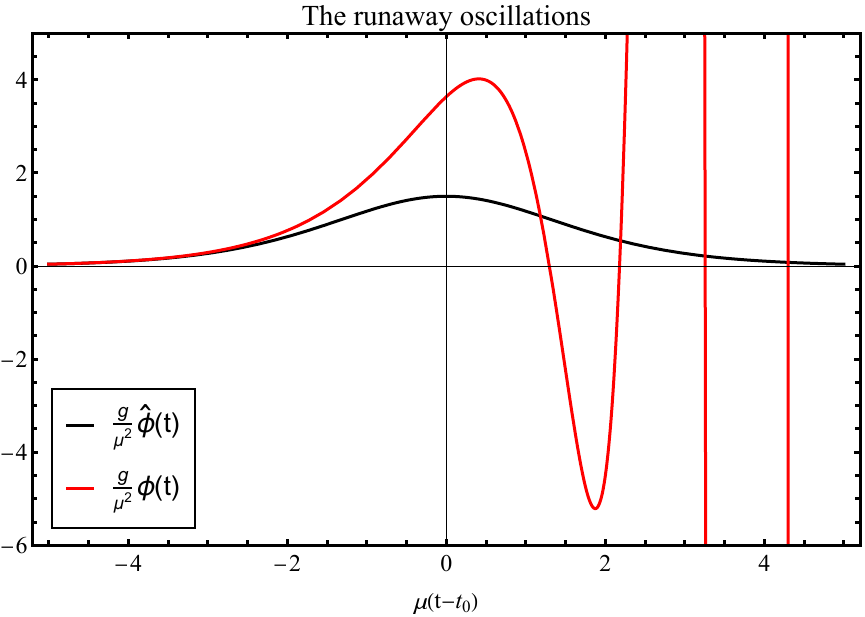}
	\caption{\label{fig:SympOSFT8} The exponential rolling tachyon solution at $\Lambda\mu^2=1/2$, shown in red, compared to the solution of the original $\phi^3$ theory, shown in black. In the effective field theory the vacuum decay enters a phase of violent oscillation. The peaks and troughs of the oscillation have been cropped out of the figure after the first two turns. In the original $\phi^3$ theory, by contrast, the configuration returns to the unstable vacuum in the infinite future.}
\end{figure}

Nevertheless, the perturbative construction of the solution proceeds without problem. Applying the stub operator to \eq{pert_exp} gives
\begin{equation}
\phi(t) = \lambda e^{\mu t} -\frac{g e^{-3\Lambda\mu^2}}{3\mu^2} \lambda^2 e^{2\mu t} + \frac{g^2e^{-8\Lambda\mu^2}}{12\mu^4}\lambda^3 e^{3\mu t} - \mathcal{O}(\lambda^4).\label{eq:roll_forward}
\end{equation}
The solution appears to exist even nonperturbatively because the expansion in powers of $e^{\mu t}$ has infinite radius of convergence. The field slowly falls off the unstable vacuum in the distant past and then transitions to a phase of violent, runaway oscillation. This is shown in figure \ref{fig:SympOSFT8}. The picture is quite analogous to rolling tachyon solutions found in $p$-adic string theory and open SFT~\cite{Moeller,Hata,Erbin2}. However, the physics appears to be incompatible with what we know about the rolling tachyon background in scalar field theory. The oscillations continue forever and never settle back to the unstable vacuum.

\begin{figure}[t]
	\centering
	\includegraphics[scale=1.5]{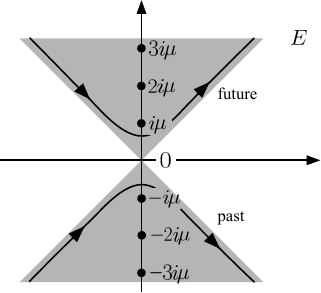}
	\caption{\label{fig:SympOSFT2} Because the rolling tachyon solution in the effective field theory diverges faster than exponentially for large real energies, to describe the time evolution we must implement a Fourier transform over a contour in the quadrant above or below the real axis. Closing the contours at infinity we obtain an infinite sum of residues at positive or respectively negative integer multiples of $i\mu$. This creates two solutions given respectively as an expansion in powers of $e^{-\mu t}$ or $e^{\mu t}$ with infinite radius of convergence. The former gives a good representation of the evolution of the background towards the future, while the later gives a better representation in the past.}
\end{figure}

There is a solution where the oscillation does settle to the unstable vacuum. This is simply the time reverse of \eq{roll_forward}. Both solutions have a common origin. Both derive from applying the stub operator to an expansion of the original $\phi^3$ solution. The solution \eq{roll_forward} derives from the expansion in exponential modes, while its time reverse comes from the expansion in inverse exponential modes. In the original $\phi^3$ theory these expansions converge to different portions of the same nonperturbative solution. In the effective field theory, however, the expansions have infinite radius of convergence and create different solutions. Another point of view comes by inspection of the Fourier transform 
\begin{equation}
\phi(t) = \int \frac{dE}{2\pi}\phi(E) e^{iEt}.
\end{equation}
Integration should be carried out over real energies, but due to the form of $\phi(E)$ in \eq{nonperturb_eff} this is not possible. We are forced make a choice. Either the contour is deformed below the real axis to the quadrant 
\begin{equation}\mathrm{Im}(E)<-|\mathrm{Re}(E)|,\end{equation} 
or above the real axis to the quadrant 
\begin{equation}\mathrm{Im}(E)>|\mathrm{Re}(E)|,\end{equation}
as shown in figure \ref{fig:SympOSFT2}. The two contours give respectively the solution \eq{roll_forward} or its time reverse. It is strange that the effective field theory has two rolling tachyon solutions when there should only be one. We believe that the correct interpretation is that the two solutions are physically equivalent, despite the very different nature of their time evolution. One solution is adapted for describing the early time behavior, and the other the late time behavior.

We can speculate whether something analogous happens in open SFT. The exponential rolling tachyon solution in SFT undergoes violent oscillations towards the future. A different solution might be needed to reveal the nature of the decay process at late times. Perhaps the solution ends at the tachyon vacuum, as suggested in \cite{Ellwood}. It is interesting to consider the pressure of the rolling tachyon background \cite{Sen2,Sen3}, which is proportional to the function  
\begin{equation}
f(t)=\frac{1}{1+e^{t-t_0}}-1.
\end{equation}
The exponential rolling tachyon solution in open SFT reconstructs $f(t)$ as an expansion in powers of $e^t$, for example though computation of the Ellwood invariant \cite{Ellwood2,Kudrna2}. This expansion however diverges at finite time. Another solution may be needed to reconstruct the pressure as an expansion in powers of $e^{-t}$. Perhaps the solution can be inferred from asymptotic analysis of the rolling tachyon background, following \cite{Erler6}. 

\section{Rolling tachyon in open string field theory}

The action of Witten's string field theory is
\begin{equation}S = -\frac{1}{g^2}\left[\frac{1}{2}\langle \Psi,Q\Psi\rangle + \frac{1}{3}\langle\Psi,\Psi*\Psi\rangle\right],\end{equation}
where $Q$ is the BRST operator, $\langle,\rangle$ is the BPZ inner product, and $*$ denotes Witten's open string star product. The dynamical string field $\Psi$ represents fluctuations of a D-brane system in bosonic string theory. The D-brane system is defined by a matter/ghost boundary conformal field theory (BCFT), and $\Psi$ is a Grassmann odd state of this BCFT at ghost number 1. Our conventions for Witten's open SFT follow \cite{Taylor2,Erler4}. We assume that the BCFT has three factors,
\begin{equation}\BCFT = \BCFT_{X^0}\otimes \BCFT_\mathrm{spatial}\otimes \BCFT_{bc},\end{equation}
consisting respectively of a noncompact timelike free boson $X^0$ subject to Neumann boundary conditions, an unspecified ``spatial" BCFT with central charge $25$, and the $bc$ ghost system of central charge $-26$. Following \cite{Cho} we may be interested in a ZZ brane in $c=1$ string theory, in which case the spatial BCFT will be a $c=25$ Liouville theory with type $(1,1)$ boundary condition~\cite{Zamolodchikov}. However, for our purposes the only relevant property of the spatial BCFT aside from its central charge is that its vacuum correlator on the disk is not zero: 
\begin{equation}\langle 1\rangle^\mathrm{spatial}_\mathrm{disk}=V.\label{eq:V}\end{equation}
We can interpret $V$ as the spatial volume of the D-brane. 

We fix Siegel gauge and consider the hyperbolic cosine rolling tachyon deformation of Sen~\cite{Sen2}. The solution has an expansion in a parameter $\lambda$
\begin{equation}
\Psi = \lambda\Psi_1 + \lambda^2\Psi_2+\lambda^3\Psi_3+\mathcal{O}(\lambda^4),
\end{equation}
where
\begin{equation}
\Psi_1 = \cosh X^0(0) c_1|0\rangle, \label{eq:Psi1}
\end{equation}
and the equations of motion in Siegel gauge imply
\begin{subequations}\label{eq:Psi2}
\begin{align}
\Psi_2 & = - \frac{b_0}{L_0}\Psi_1*\Psi_1,\\
\Psi_3 & = -\frac{b_0}{L_0}(\Psi_1*\Psi_2+\Psi_2*\Psi_1).
\end{align}
\end{subequations} 
The string field rolls up the potential and stops just short of the unstable vacuum $\Psi=0$ before turning back. We shift the origin of time so that the turnaround point occurs at $t=0$. In addition, by convention $\lambda$ is chosen so that the first hyperbolic cosine harmonic appears only at leading order in~$\lambda$. The goal in what follows is to use the symplectic form \eq{Omega} to compute the energy of the rolling tachyon solution up to order $\lambda^4$. This will be compared to boundary state results obtained by Kudrna~\cite{Kudrna} and later independently by Cho, Mazel, and Yin \cite{Cho}. 

\subsection{Symplectic structure}

The symplectic structure of Witten's SFT is
\begin{equation}
\Omega = -\frac{1}{g^2}\left[\frac{1}{2}\big\langle \delta\Psi,[Q,\sigma]\delta\Psi\big\rangle +\big\langle\Psi,[\sigma\delta\Psi,\delta\Psi]\big\rangle\right],\label{eq:Witten_Omega}
\end{equation}
where the graded commutator of string fields is computed with the open string star product. Note that $\delta\Psi$ is Grassmann even because $\Psi$ is Grassmann odd. This formula follows readily from \eq{Omega} after decomposing $Q_\Phi$ into a free and interacting part and shifting the grade so that the dynamical field is anticommuting (see e.g \cite{Erler5}). The interacting contribution is localized to finite time because the sigmoid approaches the identity operator in the infinite future, and then the commutator $[\delta\Psi,\delta\Psi]$ vanishes.\footnote{Each term in the commutator $[\delta\Psi,\delta\Psi]$ comes with a divergence from integration over time. We assume that the divergence is regularized so that the terms exactly cancel.}

The symplectic structure proposed by Witten \cite{Witten} is based on a sigmoid of the form
\begin{equation}
\sigma = \Theta\big(X^0(i,\overline{i})\big),
\end{equation}
where $\Theta(x)$ is the Heaviside step function and $X^0(i,\overline{i})$ is the timelike free boson evaluated at the open string midpoint $\xi = i$. Because the Witten vertex is formally local at the open string midpoint, the midpoint coordinate commutes through the open string star product: 
\begin{equation}X^0(i,\overline{i})\big(A*B\big) = \big(X^0(i,\overline{i})A\big)*B = A*\big(X^0(i,\overline{i})B\big).\end{equation}
Therefore Witten's choice of sigmoid is expected to satisfy
\begin{equation}[\sigma\delta\Psi,\delta\Psi] = \sigma[\delta\Psi,\delta\Psi] = 0,\end{equation}
and the interacting part of the symplectic structure drops out. This simplification is highly dependent on the special geometry of the Witten vertex. We are interested in a different construction which extends more readily to other SFTs. We consider the sigmoid 
\begin{equation}
\sigma = \int \frac{dE}{2\pi}\sigma(E) e^{iE x^0},\label{eq:sigmoid}
\end{equation}
where $x^0$ is the position zero mode operator
\begin{equation}
x^0 = \frac{1}{\pi}\int_0^\pi d\sigma X^0(e^{i\sigma},e^{\overline{i}\sigma}),
\end{equation} 
or equivalently the time coordinate of the open string center of mass. The sigmoid is independent of the spatial BCFT, and its profile in time is defined through Fourier modes $\sigma(E)$ satisfying the boundary condition \eq{sigma_om_bc}
\begin{equation}\lim_{E\to 0} \dot{\sigma}(E) = 1.\end{equation}
The open string center of mass does not commute through the star product. Therefore the interacting contribution to the symplectic structure will not vanish in our case. This makes our computation of the symplectic structure quite different from Cho, Mazel, and Yin's, which is based on Witten's construction. However the final results should not depend on how the sigmoid is chosen. A slight complication is that Witten's symplectic structure is not really well-defined. We will elaborate on this in subsection \ref{subsec:Cho}, and explain how Cho, Mazel, and Yin's remedy \cite{Cho} relates to our approach. 

The rolling tachyon is a perturbative, homogeneous, time dependent solution. Such solutions come in a two dimensional phase space differing by the size of the deformation parameter $\lambda$ and the origin of the time coordinate $t_0$. The differential on this phase space takes the form
\begin{align}
\delta\Psi & = \delta\lambda\big(\Psi_1 + 2\lambda\Psi_2 + 3\lambda^2\Psi_3+\mathcal{O}(\lambda^3)\big)\nonumber\\
&\ \ \ -\delta t_0 \big(\lambda \dot{\Psi}_1 + \lambda^2 \dot{\Psi}_2 +\lambda^3 \dot{\Psi}_3 +\mathcal{O}(\lambda^4)\big),
\end{align}
where the dot denotes the time derivative. The time derivative of a string field is defined by the action of the momentum zero mode operator
\begin{equation}
ip_0 = \oint\frac{d\xi}{2\pi i}\d X^0(\xi).
\end{equation}
Plugging $\delta\Psi$ into \eq{Witten_Omega} we find a perturbative expansion for the symplectic structure 
\begin{equation}
\Omega = \delta t_0\delta\lambda\Big[\lambda \Omega_1 + \lambda^2\Omega_2 + \lambda^3\Omega_3+\mathcal{O}(\lambda^4) \Big],
\end{equation}
with coefficients
\begin{subequations}  
\begin{align}
\Omega_1 & = \frac{1}{g^2}\langle \dot{\Psi}_1,[Q,\sigma]\Psi_1\rangle,\\
\Omega_2 & = \frac{1}{g^2}\bigg[2\langle \dot{\Psi}_1,[Q,\sigma]\Psi_2\rangle+\langle \dot{\Psi}_2,[Q,\sigma]\Psi_1\rangle+\big\langle\Psi_1,[\dot{\Psi}_1,\sigma\Psi_1]-[\sigma\dot{\Psi}_1,\Psi_1]\big\rangle\bigg],\\
\Omega_3 & =  \frac{1}{g^2}\bigg[3\langle \dot{\Psi}_1,[Q,\sigma]\Psi_3\rangle+2\langle \dot{\Psi}_2,[Q,\sigma]\Psi_2\rangle+\langle \dot{\Psi}_3,[Q,\sigma]\Psi_1\rangle\nonumber\\
&\ \ \ \ \ \ \ \ +\big\langle\Psi_2,[\dot{\Psi}_1,\sigma\Psi_1]-[\sigma\dot{\Psi}_1,\Psi_1]\big\rangle+\big\langle\Psi_1,[\dot{\Psi}_2,\sigma\Psi_1]-[\sigma\dot{\Psi}_2,\Psi_1]\big\rangle\nonumber\\
&\ \ \ \ \ \ \ \ +2\big\langle\Psi_1,[\dot{\Psi}_1,\sigma\Psi_2]-[\sigma\dot{\Psi}_1,\Psi_2]\big\rangle\bigg].
\end{align}
\end{subequations}
From this we can compute the energy
\begin{equation}
E(\lambda) = \frac{\lambda^2}{2}\Omega_1 + \frac{\lambda^3}{3}\Omega_2+ \frac{\lambda^4}{4}\Omega_3+\mathcal{O}(\lambda^5).\label{eq:E_SFT}
\end{equation}
We evaluate the coefficients $\Omega_1,\Omega_2,\Omega_3$ for the rolling tachyon in the following subsections.

\subsection{Energy to leading order $\lambda^2$}

We start with the leading order coefficient $\Omega_1$. This coefficient is somewhat trivial because it only determines the energy of a solution to the Klein-Gordon equation. However, it is important to see what this calculation looks like in the language of string field theory. 

To start we must compute the commutator of the BRST operator with the sigmoid. For this we need
\begin{equation} [Q,x^\mu] = 2\gamma^\mu,\label{eq:Qx}\end{equation}
where $\gamma^\mu$ is the zero mode of the photon vertex operator
\begin{equation}
\gamma^\mu = \oint \frac{d\xi}{2\pi i} \frac{1}{\xi} c\d X^\mu(\xi) = -\frac{i}{\sqrt{2}}\sum_{n\in\mathbb{Z}} c_n\alpha^\mu_{-n}.
\end{equation}
Furthermore
\begin{equation}
[x^\mu,\gamma^\nu] = c_0 \eta^{\mu\nu}.\label{eq:xgamma}
\end{equation}
With our choice of sigmoid \eq{sigmoid} and using \eq{Qx} we obtain
\begin{align}
[Q,\sigma]
& = \int\frac{dE}{2\pi}i \sigma(E)\int_0^E dE'  e^{i(E-E')x^0}(2\gamma^0) e^{iE'x^0}.
\end{align}
We want to commute $\gamma^0$ outside of the plane wave vertex operators. One $\gamma^0$ will be commuted to the left, and the other to the right. Using \eq{xgamma} we obtain
\begin{align}
[Q,\sigma]
& = \int\frac{dE}{2\pi}i \sigma(E)\int_0^E dE'\Big[\Big(\gamma^0 e^{iE x^0} -i(E-E')c_0 e^{iE x^0}\Big) +\Big(e^{iE x^0}\gamma^0 +iE' e^{iE x^0} c_0\Big) \Big]\nonumber\\
& = \int\frac{dE}{2\pi}i \sigma(E)\int_0^E dE'\Big[\gamma^0 e^{iE x^0}+e^{iE x^0}\gamma^0  -i(E-2E')c_0 e^{iE x^0}\Big].
\end{align}
Integrating $E'$, the last term drops out and the first two terms are multiplied by $E$. This computes the time derivative of the sigmoid, so we finally obtain
\begin{equation}
[Q,\sigma] = \gamma^0 \dot{\sigma} +\dot{\sigma}\gamma^0,\label{eq:Qsigma}
\end{equation}
where 
\begin{equation}\dot{\sigma} = \int \frac{dE}{2\pi} \dot{\sigma}(E)e^{iE x^0}.\label{eq:sigmadot}\end{equation}
The time derivative makes it clear that the symplectic structure is localized to finite time.

We introduce energy eigenstates
\begin{equation} 
|E\rangle = e^{iE X^0(0,0)}|0\rangle,\ \ \ \ \langle E| = |E\rangle^\star,
\end{equation}
where $\star$ denotes BPZ conjugation. These satisfy
\begin{equation}p_0 |E\rangle = E|E\rangle,\ \ \ \langle E|p_0 = -E\langle E|,\end{equation}
and are normalized 
\begin{equation}
\langle E|c_{-1}c_0 c_1|E'\rangle = 2\pi V\delta(E+E').
\end{equation}
The dual eigenstate carries eigenvalue $-E$ because the momentum operator is BPZ odd. With this we can write the leading order solution as
\begin{equation}
\Psi_1 = \frac{1}{2}c_1\Big(|i\rangle +|-i\rangle\Big).
\end{equation}
In addition
\begin{equation}
(\dot{\Psi}_1)^\star = -\frac{1}{2}\Big(\langle i|-\langle -i|\Big)c_{-1}.
\end{equation}
The coefficient $\Omega_1$ can then be found by computing the contraction 
\begin{equation}
\Omega_1 =  -\frac{1}{4g^2}\Big(\langle i|-\langle -i|\Big)c_{-1}\Big(\gamma^0 \dot{\sigma}+\dot{\sigma}\gamma^0\Big)c_1\Big(|i\rangle +|-i\rangle\Big).
\end{equation}
The oscillator part of $\gamma^0$ drops out because lowering operators annihilate the vacuum. Using
\begin{equation}
\gamma^0 = i c_0 p_0 + \mathrm{oscillator\ part},
\end{equation}
we obtain
\begin{align}
\Omega_1 & =  -\frac{1}{4g^2}\bigg[ \Big(\langle i|+\langle -i|\Big)c_{-1}c_0 c_1 \dot{\sigma}\Big(|i\rangle+|-i\rangle\Big)\nonumber\\ &\ \ \ \ \ \ \ \ \ \ \ \ \ 
 -\Big(\langle i|-\langle -i|\Big)c_{-1}c_0 c_1 \dot{\sigma}\Big(|i\rangle-|-i\rangle\Big)\bigg].
\end{align}
Substituting \eq{sigmadot}
\begin{align} 
\Omega_1 & = -\frac{1}{4 g^2}\int \frac{dE}{2\pi} \dot{\sigma}(E) \bigg[ \Big(\langle i|+\langle -i|\Big)c_{-1}c_0 c_1 \Big(|i+E\rangle+|-i+E\rangle\Big)\nonumber\\ &\ \ \ \ \ \ \ \ \ \ \ \ \ \ \ \ \ \ \ \ \ \ \ \ \ \ \ \ \ \  
 -\Big(\langle i|-\langle -i|\Big)c_{-1}c_0 c_1\Big(|i+E\rangle-|-i+E\rangle\Big)\bigg].
\end{align}
Contracting the states and dual states we obtain delta functions which equate the energy to one of three values:
\begin{equation}-2i,\ \ \ 0,\ \ \ 2i.\end{equation}
Tallying the cross terms contributing at each energy we obtain
\begin{align} 
\Omega_1 = -\frac{V}{4 g^2}\int dE &\Big[ \,\dot{\sigma}(-2i)\delta(E+2i)\big(1-1\big)\nonumber\\
& +\dot{\sigma}(0)\delta(E)\big(1+1+1+1\big)\nonumber\\
& +\dot{\sigma}(2i)\delta(E-2i)\big(1-1\big)\Big].
\end{align}
As expected, the contributions from the sigmoid at nonzero energy cancel out. Using $\dot{\sigma}(0)=1$ we obtain
\begin{equation}\Omega_1= -\frac{V}{g^2}.\end{equation}
Secretly this calculation is the same as the one given in subsection \ref{subsec:eff_leading}. The result agrees with \eq{eff_Omega1_result} keeping in mind the $1/g^2$ normalization of the open SFT action and that in $\alpha'=1$ units the tachyon mass$^2$ is $-1$.

\subsection{Energy to order $\lambda^4$}

To determine the higher order coefficients we will repeatedly assume, without explicitly demonstrating, that contributions from the sigmoid at nonzero energy cancel out.\footnote{We checked this for $\Omega_2$, but did not for $\Omega_3$.} In this way we can argue that $\Omega_2$ vanishes, as explained  in subsection \ref{subsec:eff_leading}. Therefore we focus on $\Omega_3$. The goal of this subsection is to reduce $\Omega_3$ to correlation functions on the surfaces of the 4-point amplitude in Siegel gauge. The evaluation of these correlation functions will be undertaken in the next subsection.

The coefficient $\Omega_3$ has contributions from the overlap of $\Psi_1$ and $\Psi_3$, for example
\begin{equation}
\langle \dot{\Psi}_1,[Q,\sigma]\Psi_3\rangle.\label{eq:Psi1Psi3}
\end{equation}
We observe that the leading order solution contains no creation operators. Decomposing
\begin{equation}L_0 = p^2-1+N,\end{equation}
where $N$ is the mode counting operator, this means that $\Psi_1$ is annihilated by $N$. We can write this equivalently as 
\begin{equation}\delta_N\Psi_1=\Psi_1,\end{equation}
where $\delta_N$ is the projector onto the kernel of $N$. With our choice of sigmoid, $[Q,\sigma]$ commutes through the mode counting operator. This means that \eq{Psi1Psi3} can be written
\begin{equation}
\langle \dot{\Psi}_1,[Q,\sigma]\delta_N\Psi_3\rangle.
\end{equation}
Next we assume that only contributions from the sigmoid at zero energy can contribute. Since $\Psi_1$ comes with energy $E=\pm i$, we only have to worry about contributions from $\Psi_3$ with energy $E=\pm i$. Thus
\begin{equation}
\langle \dot{\Psi}_1,[Q,\sigma]\Psi_3\rangle = 
\langle \dot{\Psi}_1,[Q,\sigma]\delta_N\delta_{E^2+1}\Psi_3\rangle+\text{nonzero energy},
\end{equation}
where $\delta_{E^2+1}$ is the projector onto states with $E^2+1=0$. Next we observe that if both $N$ and $E^2+1$ have to vanish, the full dilatation generator $L_0$ must vanish. Therefore 
\begin{equation}
\langle \dot{\Psi}_1,[Q,\sigma]\Psi_3\rangle = 
\langle \dot{\Psi}_1,[Q,\sigma]\delta_N\delta_{L_0}\Psi_3\rangle+\text{nonzero energy},
\end{equation}
where $\delta_{L_0}$ is the projector onto the kernel of $L_0$. Because the hyperbolic cosine deformation is exactly marginal we know that 
\begin{equation}\delta_{L_0}\Psi_3=0.\end{equation}
Therefore the $\Psi_1$-$\Psi_3$ contribution to $\Omega_3$ can only give terms with the sigmoid evaluated at nonzero energy. Since we know that all such terms must cancel, we can ignore $\Psi_1$-$\Psi_3$ contributions.

Next consider the contribution from a pair of $\Psi_2$s. Plugging in \eq{Psi2} we have
\begin{equation}
\langle \dot{\Psi}_2,[Q,\sigma]\Psi_2\rangle = \left\langle[\dot{\Psi}_1,\Psi_1],\frac{b_0}{L_0}[Q,\sigma]\frac{b_0}{L_0}\Psi_1^2\right\rangle,
\end{equation}
where we suppress the $*$ in Witten's star product. Note that
\begin{equation}
\frac{b_0}{L_0}[Q,\sigma]\frac{b_0}{L_0} = \frac{1}{L_0}[L_0,\sigma]\frac{b_0}{L_0}= \left[\sigma,\frac{b_0}{L_0}\right],
\end{equation}
where in canceling $L_0$ we note that states in the kernel of $L_0$ are not present. Therefore this contribution contains only one propagator. The remaining parts of $\Omega_3$ involve a pair of $\Psi_1$s and a $\Psi_2$ and also contain a single propagator. Up to now we have expressed $\Omega_3$ as
\begin{align}
\Omega_3 & = \frac{1}{g^2}\Bigg[2\left\langle[\dot{\Psi}_1,\Psi_1],\left[\sigma,\frac{b_0}{L_0}\right]\Psi_1^2\right\rangle-2\left\langle\Psi_1,\left[\sigma\frac{b_0}{L_0}\Psi_1^2,\dot{\Psi}_1\right]-\left[\frac{b_0}{L_0}\Psi_1^2,\sigma\dot{\Psi}_1\right]\right\rangle\nonumber\\
&\ \ \ \ \ \ \ \ -\left\langle \Psi_1,\left[\sigma\Psi_1,\frac{b_0}{L_0}[\dot{\Psi}_1,\Psi_1]\right]-\left[\Psi_1,\sigma\frac{b_0}{L_0}[\dot{\Psi}_1,\Psi_1]\right]\right\rangle \nonumber\\
&\ \ \ \ \ \ \ \ +\left\langle[\sigma\dot{\Psi}_1,\Psi_1]-[\dot{\Psi}_1,\sigma\Psi_1],\frac{b_0}{L_0}\Psi_1^2\right\rangle\Bigg]+\text{nonzero energy}.\label{eq:Omega3_1}
\end{align}
All terms involve star products of two Fock space states connected by a Siegel gauge propagator. Therefore they may be evaluated as correlation functions on the same kind of surface which appears in the four point amplitude in Siegel gauge.

To further simplify we want to write $\Omega_3$ more explicitly as an overlap of the form
\begin{equation}\left\langle \text{(Fock)}^2,\frac{b_0}{L_0}\text{(Fock)}^2\right\rangle.\end{equation}
Doing this would be fairly easy if we could make use of cyclicity of the open string star product. But we need to be careful because cyclicity can generate hidden boundary terms. To deal with this we make use of tau regularization \cite{CovL}. Note that \eq{Omega3_1} has been written so that all terms are localized to finite time. Therefore we can insert the operator $\tau$ in front of all instances of $\dot{\Psi}_1$ without changing the result. The factor of $\tau$ regularizes volume divergences so that we can freely use cyclicity. In this way we can manipulate $\Omega_3$ to the form
\begin{align}
\Omega_3 & = \frac{1}{g^2}\Bigg[2\left\langle[\tau\dot{\Psi}_1,\Psi_1],\left[\sigma,\frac{b_0}{L_0}\right]\Psi_1^2\right\rangle-\left\langle 2\sigma[\tau\dot{\Psi}_1,\Psi_1]-3[\sigma\tau\dot{\Psi}_1,\Psi_1]+[\tau\dot{\Psi}_1,\sigma\Psi_1],\frac{b_0}{L_0}\Psi_1^2\right\rangle\nonumber\\
&\ \ \ \ \ \ \ \ +\left\langle[\tau\dot{\Psi}_1,\Psi_1],\frac{b_0}{L_0}\Big(2\sigma\Psi_1^2 -[\sigma\Psi_1,\Psi_1]\Big)\right\rangle\Bigg]+\text{nonzero energy}.
\end{align}
The terms are again localized to finite time so we can drop the $\tau$. It is helpful to define products
\begin{subequations}
\begin{align}
\sigma_+(A,B)& =2\sigma[A,B] -[\sigma A,B] -[A,\sigma B],\\
\sigma_-(A,B)& = [\sigma A,B]-[A,\sigma B],
\end{align}
\end{subequations}
which are respectively graded antisymmetric and graded symmetric in $A$ and $B$. Both products are localized to finite time. In terms of these we can write
\begin{align}
\Omega_3 & = \frac{1}{g^2}\Bigg[2\left\langle[\dot{\Psi}_1,\Psi_1],\left[\sigma,\frac{b_0}{L_0}\right]\Psi_1^2\right\rangle-\left\langle\sigma_+(\dot{\Psi}_1,\Psi_1),\frac{b_0}{L_0}\Psi_1^2\right\rangle+2\left\langle\sigma_-(\dot{\Psi}_1,\Psi_1),\frac{b_0}{L_0}\Psi_1^2\right\rangle\nonumber\\
&\ \ \ \ \ \ \ \ \ +\frac{1}{2}\left\langle[\dot{\Psi}_1,\Psi_1],\frac{b_0}{L_0}\sigma_+(\Psi_1,\Psi_1)\right\rangle\Bigg]+\text{nonzero energy}.\label{eq:Omega3_2}
\end{align}
To arrive at the final expression we want to drop nonzero energy contributions wherever possible. For this we decompose into exponential modes
\begin{equation}\Psi_1 = \frac{1}{2}\Big(\Psi_++\Psi_-\Big),\end{equation}
where
\begin{equation}\Psi_\pm = c e^{\pm X^0(0,0)}|0\rangle.\end{equation}
Note that
\begin{subequations}
\begin{align}
[\dot{\Psi}_1,\Psi_1] & = \frac{1}{2}\Big(\Psi_+^2-\Psi_-^2\Big),\\
\sigma_+(\dot{\Psi}_1,\Psi_1) & = \frac{1}{4}\Big(\sigma_+(\Psi_+,\Psi_+)- \sigma_+(\Psi_-,\Psi_-)\Big),\\
\sigma_-(\dot{\Psi}_1,\Psi_1) &= \frac{1}{2}\sigma_-(\Psi_+,\Psi_-).
\end{align}
\end{subequations}
Plugging this in to \eq{Omega3_2} we finally arrive at
\begin{align}
\Omega_3 & = \frac{1}{g^2}\Bigg[\frac{1}{2}\left\langle \Psi_+^2,\left[\sigma,\frac{b_0}{L_0}\right]\Psi_-^2\right\rangle +\frac{1}{4}\left\langle \sigma_-(\Psi_+,\Psi_-),\frac{b_0}{L_0}[\Psi_+,\Psi_-]\right\rangle\nonumber\\
&\ \ \ \ \ \ \ \ +\frac{1}{8}\left\langle  \sigma_+(\Psi_-,\Psi_-),\frac{b_0}{L_0}\Psi_+^2\right\rangle - \frac{1}{8}\left\langle \sigma_+(\Psi_+,\Psi_+),\frac{b_0}{L_0}\Psi_-^2\right\rangle\Bigg].\label{eq:Omega3}
\end{align} 
All terms containing the sigmoid evaluated at nonzero energy have been dropped. 

\subsection{Oscillator expectation values}

To evaluate these correlators we need to uniformize the geometry of the 4-point amplitude in Siegel gauge. This can be done analytically \cite{Taylor,Giddings,Sloan,Samuel}, but the formulas are cumbersome. To get numbers it is more straightforward to do a numerical computation. Cho, Mazel, and Yin evaluate the correlation functions using Virasoro conservation laws and level truncation \cite{Rastelli}. We will proceed in a different way using squeezed state oscillator methods developed by Taylor~\cite{Taylor,Taylor3,Taylor4,Coletti,Ellwood3}. This approach approximates the correlation function by imposing an upper limit on the mode number of Neumann coefficients appearing in the squeezed state representation of the Witten vertex. The technique is more efficient than level truncation because it only requires dealing with matrices whose size grows linearly with mode number, rather than fields whose number grows exponentially with level. The implementation is also more accessible. The main limitation is that the method only applies to free boson CFTs. We will assume that the spatial BCFT consists of 25 noncompact free bosons subject to Neumann boundary conditions. However, since the rolling tachyon solution does not excite primaries in the spatial BCFT, the results should not depend on this assumption. 

The three string vertex $|V_3\rangle$ lives in the tensor product of three copies of the BCFT vector space 
and satisfies
\begin{equation}
\langle A,B*C\rangle = A^\star\otimes B^\star\otimes C^\star|V_3\rangle,
\end{equation}
where $A^\star,B^\star$ and $C^*$ are the BPZ conjugates of the states $A,B$ and $C$. The BPZ conjugate vertex $\langle V_3|$ satisfies
\begin{equation}
\langle A* B,C\rangle = \langle V_3|A\otimes B\otimes C.
\end{equation}
Below we give an expression for the vertex as a squeezed state in matter and ghost oscillators on a D25-brane in flat Minkowski space. This expression is is sometimes called the oscillator vertex, and takes the form
\begin{align}
|V_3\rangle & = \kappa \int \frac{d^{26}k^1d^{26}k^2d^{26}k^3}{(2\pi)^{3\cdot 26}}(2\pi)^{26} \delta^{26}(k^{123})\exp\Bigg[-\frac{1}{2}a^A_{-m}\cdot N_{mn}^{AB} a^B_{-n}-a^A_{-m}\cdot N_{m0}^{AB}k^B\nonumber\\
&\ \ \ \ \ \ \ \ \ \ \ \ \ \ \  -\frac{1}{2}k^A\cdot N_{00}^{AB}k^B -c_{-m}^A X_{mn}^{AB}b_{-n}^B - c_{-m}^A X^{AB}_{m0}b_0\Bigg]|+,k^1\rangle\otimes|+,k^2\rangle\otimes|+,k^3\rangle.\label{eq:V3}
\end{align}
The dual vertex is 
\begin{align}
\langle V_3| & = \kappa \int \frac{d^{26}k^1d^{26}k^2d^{26}k^3}{(2\pi)^{3\cdot 26}V^3}(2\pi)^{26} \delta^{26}(k^{123})\langle +,k^1|\otimes \langle +,k^2|\otimes \langle+,k^3|\exp\Bigg[-\frac{1}{2}a^A_m\cdot \overline{N}_{mn}^{AB} a^B_{n}\nonumber\\
&\ \ \ \ \ \ \ \ \ \ \ \ \ \ \  -a^A_{m}\cdot \overline{N}_{m0}^{AB}k^B-\frac{1}{2}k^A\cdot \overline{N}_{00}^{AB}k^B -c_{m}^A \overline{X}_{mn}^{AB}b_{n}^B - c_{m}^A \overline{X}^{AB}_{m0}b_0\Bigg].
\end{align}
Let us explain the notation. The squeezed state is built from Fock vacua
\begin{subequations}
\begin{align}
|-,k\rangle & = c_1 e^{i k\cdot X(0,0)}|0\rangle,\ \ \ \ \ \ \ \ \ \langle -,k| = |-,k\rangle^\star,\\
|+,k\rangle & = c_0c_1 e^{ik\cdot X(0,0)}|0\rangle,\ \ \ \ \ \ \ \langle +,k| = |+,k\rangle^\star.
\end{align}
\end{subequations}
which are momentum eigenstates
\begin{equation}
p_\mu |\pm,k\rangle = k_\mu|\pm,k\rangle,\ \ \ \ \ \langle \pm,k|p_\mu = -k_\mu \langle\pm,k|,
\end{equation}
normalized as
\begin{equation}
\langle +,k|-,k'\rangle = \langle -,k|+,k'\rangle = (2\pi)^{26}\delta^{26}(k+k').
\end{equation}
We introduce normalized matter oscillators
\begin{equation}a^{\mu}_m = \frac{1}{\sqrt{|m|}}\alpha^\mu_m,\ \ \ \ (m\neq 0),\end{equation}
which satisfy
\begin{equation}[a_m^\mu, a_{-n}^\nu] = \delta_{mn}\eta^{\mu\nu}.\end{equation}
All oscillators with positive mode index annihilate the Fock vacua
\begin{equation}a_m^\mu|\pm,k\rangle = b_m|\pm,k\rangle=c_m|\pm,k\rangle,\ \ \ \ (m\geq 1),\end{equation}
and with negative mode index act as creation operators. When possible Lorentz indices are left implicit with the notation
\begin{equation}u\cdot v = u^\mu v^\nu\eta_{\mu\nu},\end{equation}
The oscillators carry a string label $A,B$ that indicates which copy of the BCFT vector space they belong to. Typically $A,B$ are assumed to take values $1,2,3$ corresponding to the three Fock vacua of the oscillator vertex counted left to right. However in more general Feynman graphs $A,B$ may label any of several string states that appear in the graph. Repeated mode indices are summed over positive integers and repeated string indices summed over all string labels. Finally, the numbers
\begin{align}
& N_{mn}^{AB},\ \ \ \ N_{m0}^{AB},\ \ \ \ N_{00}^{AB}, \ \ \ \ X_{mn}^{AB},\ \ \ \ X_{m0}^{AB}, \nonumber\\
& \overline{N}_{mn}^{AB},\ \ \ \ \overline{N}_{m0}^{AB},\ \ \ \ \overline{N}_{00}^{AB},\ \ \ \ \overline{X}_{mn}^{AB},\ \ \ \ \overline{X}_{m0}^{AB}
\end{align}
are Neumann coefficients. Formulas were given by Gross and Jevicki~\cite{Gross1,Gross2}, which can be found conveniently displayed in \cite{Taylor2}.\footnote{The ghost Neumann coefficient $X^{rr}_{mn}$ in equation (195) of \cite{Taylor2} is missing a factor of $m$.}  Care must be taken in extracting these numbers from the literature due slight differences in convention and sometimes errors. Our notation and conventions for the Neumann coefficients follows \cite{Taylor}, and our implementation can be found in the {\tt Mathematica} file accompanying this document. We have checked our formulas by reproducing Taylor's results for the off-shell 4 tachyon amplitude in Siegel gauge \cite{Taylor}. The string indices $A,B$ on the  Neumann coefficients are always mapped to values $1,2,3$, regardless of how the strings are labeled on the larger diagram. They have the cyclicity property
\begin{equation}N_{mn}^{A+1,B+1} = N_{mn}^{AB},\ \ \ \ X_{mn}^{A+1,B+1}=X_{mn}^{AB},\end{equation}
where the increment in the string label is taken mod 3 and $m,n$ can be zero. The Neumann coefficients in the BPZ conjugate vertex are
\begin{align}& \overline{N}_{mn}^{AB} = (CN^{AB}C)_{mn},\ \ \ \ \overline{N}_{m0}^{AB} = - C_{mn}N_{n0}^{AB},\ \ \ \ \overline{N}^{AB}_{00}=N^{AB}_{00} = \delta^{AB}\ln\left(\frac{27}{16}\right)\nonumber\\
& \overline{X}_{mn}^{AB} = (CX^{AB}C)_{mn},\ \ \ \ \overline{X}_{m0}^{AB} = C_{mn}X_{n0}^{AB},\end{align}
where $C_{mn}=(-1)^m\delta_{mn}$ is called the twist matrix. The nonzero mode Neumann coefficients may be regarded as infinite dimensional matrices in the mode label. Products of such matrices will be written as 
\begin{equation}(AB)_{mn}= A_{mk}B_{kn}\end{equation}
to reduce index clutter.

\begin{figure}[t]
	\centering
	\includegraphics[scale=1]{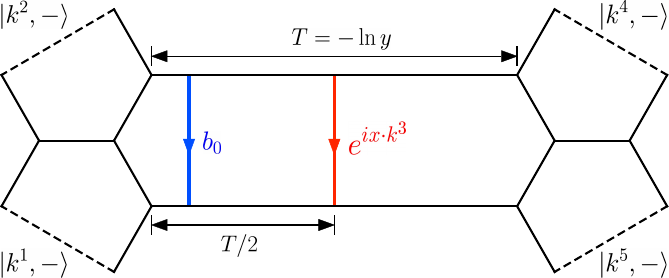}
	\caption{\label{fig:SympOSFT3} The correlation function \eq{4tach} is defined by connecting the star product of two Fock vacua to the star product of another two Fock vacua by a Siegel gauge propagator strip. A plane wave insertion of momentum appears exactly half way along the length of the propagator strip.}
\end{figure} 

\begin{figure}[t]
	\centering
	\includegraphics[scale=.7]{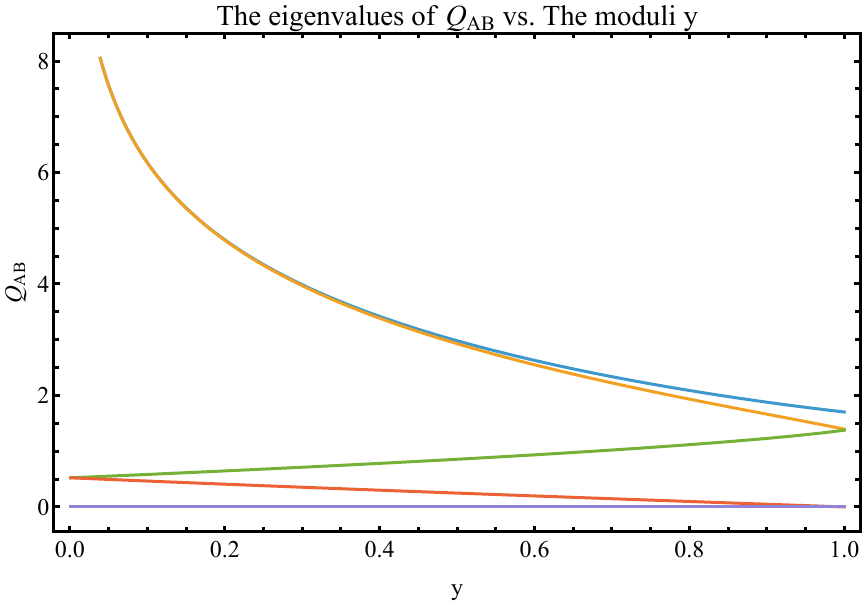}
	\caption{\label{fig:SympOSFT7} Eigenvalues of the quadratic form $Q^{AB}$ plotted as a function of the modulus $y$. The results are obtained by truncating the matrices at mode number $L=100$. The zero eigenvalue corresponds to the momentum $k^3$ inserted in the propagator, and the remaining eigenvalues are positive. }
\end{figure}

Every term in $\Omega_3$ can be reduced to a correlation function of the form
\begin{equation}
\bigg\langle y^{L_0/2}\Big(c_1|k^1\rangle*c_1|k^2\rangle\Big), b_0e^{ik^3\cdot x}~ y^{L_0/2}\Big(c_1|k^4\rangle*c_1|k^5\rangle\Big)\bigg\rangle = (2\pi)^{26} \delta^{26}(k^{12345})F(k^1,k^2,k^3,k^4,k^5),\label{eq:4tach}
\end{equation} 
where $k^1,...,k^5$ are five momenta and $y$ is a modulus. By factoring out a momentum conserving delta function we define a function $F(k^1,...,k^5)$ whose dependence on the modulus $y$ will be kept implicit. The sum of the arguments of $F(k^1,...,k^5)$ is assumed to be zero by momentum conservation. The correlation function is almost the integrand of the 4-point amplitude of off-shell tachyons in Siegel gauge. The difference comes from the plane wave insertion of momentum on the propagator strip. The insertion appears exactly half-way along the propagator strip of length $T=-\ln y$. The geometry is illustrated in figure \ref{fig:SympOSFT3}.  Using the oscillator vertex we can write the correlation function as 
\begin{align}
(2\pi)^{26} &\delta^{26}(k^{12345}) F(k^1,k^2,k^3,k^4,k^5)  \nonumber\\
& = \Big(\langle V_3||-,k^1\rangle\otimes |-,k^2\rangle \otimes y^{L_0/2}\Big)b_0e^{ik^3\cdot x}\Big(y^{L_0/2}\otimes \langle -,k^4|\otimes \langle -,k^5||V_3\rangle\Big). \label{eq:osc4tach}
\end{align}
The right hand side can be evaluated using squeezed state contraction formulas. The first formula we need is
\begin{align}
& \langle 0|\exp\left[-u_i a_i -\frac{1}{2} a_i A_{ij} a_j\right]\exp\left[-v_i a_i^\dag -\frac{1}{2} a_i^\dag B_{ij} a_j^\dag\right]|0\rangle\nonumber\\
& \ =\frac{1}{\sqrt{\det(1-AB)}}\exp\Bigg[-\frac{1}{2}u_i\left(B\frac{1}{1-AB}\right)_{ij}u_j+v_i\left(\frac{1}{1-AB}\right)_{ij}u_j - \frac{1}{2}v_i\left(\frac{1}{1-AB}A\right)_{ij}v_j\Bigg],
\end{align} 
where  $a_i,a_i^\dag$ are Grassmann even creation/annihilation operators satisfying 
\begin{align}
[a_i,a_j^\dag] = \delta_{ij},\ \ \ \ \ a_i|0\rangle = 0,\ \ \ \langle 0| a_i^\dag  = 0,
\end{align} 
with unwritten commutators zero. The second formula is 
\begin{align}
\langle 0|\exp\Big(-b_i X_{ij} c_j\Big) \exp\Big(-c^\dag_i Y_{ij}b^\dag_j\Big)|0\rangle = \det(1-XY),
\end{align}
where $b_i,c_i,b_i^\dag,c_i^\dag$ are Grassmann odd creation/annihilation operators satisfying
\begin{align}
[b_i,c_j^\dag]  = [c_i,b_j^\dag] = \delta_{ij},\ \ \ \ \ b_i|0\rangle = c_i|0\rangle = 0,\ \ \ \langle 0|b_i^\dag = \langle 0|c_i^\dag = 0,
\end{align}
with unwritten commutators zero. Applying to \eq{osc4tach} gives the result
\begin{equation}
F(k^1,k^2,k^3,k^4,k^5)=\frac{\kappa^2 }{y}\frac{\det\Big(1-\big(C\hat{X}^{11}(y,y)\big)^2\Big)}{\Big[\!\det\Big(1-\big(C\hat{N}^{11}(y,y)\big)^2\Big)\!\Big]^{13}}\exp\left[-\frac{1}{2}k^A\cdot Q^{AB}k^B\right].\label{eq:4tach_matrix}
\end{equation}
This is a Gaussian function of the momenta $k^A$ with $A=1,2,3,4,5$. The parameters of the Gaussian are defined through products and determinants of the Neumann coefficients. We introduce 
\begin{equation}
\hat{N}^{AB}_{mn}(x,y) = x^{m/2} N^{AB}_{mn} y^{n/2},\ \ \ \ \ \hat{X}^{AB}_{mn}(x,y) = x^{m/2} X^{AB}_{mn}y^{n/2}. \label{eq:modified_NC}
\end{equation}
The quadratic form $Q^{AB}$ is given as a sum of three terms
\begin{equation}Q^{AB}= Q^{AB}_1+ Q^{AB}_2+Q^{AB}_3.\end{equation}
The first two terms are
\begin{align}
Q_1^{AB} = \ln\left(\frac{27}{16}\right)\left(\begin{matrix}2 & 1 & & & \\ 1 & 2 & & & \\ & & 0 & & \\ & & & 2 & 1 \\ & & & 1 & 2
\end{matrix}\right)^{AB},\ \ \ \ \ 
Q_2^{AB} = -\ln y\left(\begin{matrix}1 & 1 & & & \\ 1 & 1 & & & \\ & & 0 & & \\ & & & 1 & 1 \\ & & & 1 & 1
\end{matrix}\right)^{AB},
\end{align}
where unwritten matrix entries are zero. The final term is
\begin{align}
& Q_3^{AB} = \nonumber\\
& \left[\left(\begin{matrix}
\hat{N}^{12}_{m0}(y,1)-\hat{N}_{m0}^{11}(y,1) \ \ \ \ \ \   0 \ \ \ \ \ \ \ \ \\[.25cm]
\hat{N}^{13}_{m0}(y,1)-\hat{N}_{m0}^{11}(y,1)  \ \ \ \ \ \  0 \ \ \ \ \ \ \ \ \\[.25cm]
0 \ \ \ \ \ \ \ \ \ \ \ \ \ \ \ \ \ \ \ \ \ \ \ \ \ 0  \\[.25cm]
\ \ \ \ \ \ \ \ 0 \ \ \ \ \ \   \hat{N}^{12}_{m0}(y,1)-\hat{N}_{m0}^{11}(y,1) \\[.25cm]
\ \ \ \ \ \ \ \ 0 \ \ \ \ \ \   \hat{N}^{12}_{m0}(y,1)-\hat{N}_{m0}^{11}(y,1) \\[.25cm]
\end{matrix}\right)
\left(\begin{matrix}\displaystyle{\left[\frac{C\hat{N}^{11}(y,y)}{1-(C\hat{N}^{11}(y,y))^2}C\right]_{mn}} 
 \displaystyle{\left[\frac{1}{1-(C\hat{N}^{11}(y,y))^2}C\right]_{mn}} \\[.75cm] 
\displaystyle{\left[\frac{1}{1-(C\hat{N}^{11}(y,y))^2}C\right]_{mn}}
\displaystyle{\left[\frac{C\hat{N}^{11}(y,y)}{1-(C\hat{N}^{11}(y,y))^2}C\right]_{mn}}\end{matrix}\right)\right.\nonumber\\
&\left.\times\!\!\left(\begin{matrix}\!\hat{N}^{12}_{n0}(y,1)\! -\! \hat{N}^{11}_{n0}(y,1)\! & \!\hat{N}^{13}_{n0}(y,1)\! -\! \hat{N}^{11}_{n0}(y,1)\! & 0 & 0 & 0 \\[.25cm] 0 & 0 & 0 & \!\hat{N}^{12}_{n0}(y,1)\! -\! \hat{N}^{11}_{n0}(y,1)\! & \!\hat{N}^{13}_{n0}(y,1)\! -\! \hat{N}^{11}_{n0}(y,1) \! \end{matrix}\right)\!\right]^{AB}.\label{eq:QAB}
\end{align}
When $k^3=0$ this can be compared to the integrand of the 4 tachyon amplitude in Siegel gauge as written in equation (2.24) of \cite{Taylor}. Actually, because all of the relevant matrices vanish in their middle entry, the correlation function only depends on the momentum $k^3$ through the momentum conserving delta function. To make use of these formulas we truncate the matrices to finite size by placing an upper limit $L$ on mode number. Products and determinants of the truncated matrices are then evaluated numerically. It is important to mention that the quadratic form $Q^{AB}$ has no negative eigenvalues. We have confirmed this numerically, and the results are shown in figure~\ref{fig:SympOSFT7}. This implies that the correlation function vanishes for large spatial momenta but grows as an inverted Gaussian at high energies. This fits with the general observed behavior of SFT vertices and manifests transgressive locality. 

The next step is to express $\Omega_3$ in terms of this correlation function. In \eq{Omega3} $\Omega_3$ is given as a sum of four terms, and we start with the contribution
\begin{equation}
I_1 = \frac{1}{V}\left\langle \Psi_+^2,\left[\sigma,\frac{b_0}{L_0}\right]\Psi_-^2\right\rangle.\label{eq:I1}
\end{equation}
Since in the oscillator vertex \eq{V3} the spatial BCFT corresponds to that of a D25-brane in flat space, the vacuum normalization is
\begin{equation}V = (2\pi)^{25}\delta^{25}(0).\end{equation}
We express the propagator using the Schwinger representation
\begin{equation} \frac{b_0}{L_0} =b_0\int_0^1\frac{dy}{y} y^{L_0}.\label{eq:Sch_prop}\end{equation}
This is valid assuming that no divergence appears towards the $y=0$ boundary of moduli space. Divergence will be absent if the propagator acts on states with strictly positive conformal weight. Because $\Psi_-^2$ carries momentum $2i$, the minimum conformal weight which appears in the star product is 
\begin{equation}L_0 = p^2+N-1 = -(2i)^2-1 = 3.\end{equation}
Since this is positive there is no problem with the Schwinger representation of the propagator. To make use of $F(k^1,...,k^5)$ we need to move the position zero mode operator in $\sigma$ to the center of the Siegel gauge propagator strip. This can be achieved with the following manipulation: 
\begin{align}
\left[\sigma,\frac{b_0}{L_0}\right] & = b_0\int \frac{dE}{2\pi}\sigma(E)\int_0^1 \frac{dy}{y} [e^{i E x^0},y^{L_0}] \nonumber \\ & = b_0\int \frac{dE}{2\pi}\sigma(E)\int_0^1 \frac{dy}{y} y^{L_0/2}\Big(e^{i E (y^{-L_0/2}x^0y^{L_0/2})}-e^{i E (y^{L_0/2}x^0y^{-L_0/2})}\Big) y^{L_0/2}.
\end{align}
Plug in
\begin{align}
y^{L_0/2}x^0 y^{-L_0/2}& = x^0 +\frac{1}{2}\ln y [L_0,x^0] + \frac{1}{2!}\left(\frac{1}{2}\ln y\right)^2[L_0,[L_0,x^0]]+\cdots\nonumber\\
& = x^0+ i p_0 \ln y ,
\end{align}
to get
\begin{equation}
\left[\sigma,\frac{b_0}{L_0}\right] = b_0\int \frac{dE}{2\pi}\sigma(E)\int_0^1 \frac{dy}{y} y^{L_0/2}\Big(e^{i E (x^0 - i p_0 \ln y)}-e^{i E (x^0+ i p_0 \ln y )}\Big) y^{L_0/2}.
\end{equation}
We factor the exponentials using the Campbell-Baker-Hausdorff formula
\begin{equation}
e^{i E (x^0 - i p_0 \ln y)} = e^{iE x^0} e^{E(p_0 +\frac{1}{2}E)\ln y}.
\end{equation}
The result is
\begin{equation}
\left[\sigma,\frac{b_0}{L_0}\right] = b_0 \int \frac{dE}{2\pi}\dot{\sigma}(E)\int_0^1 \frac{dy}{y} y^{L_0/2}e^{iE x^0} y^{L_0/2}\left(\frac{2\sinh\big((p_0+\frac{1}{2}E)E\ln y\big)}{iE}\right).
\end{equation}
The plane wave operator is now inserted in the middle of the propagator strip. The momentum zero mode operator appears inside the hyperbolic sine which has been factored to the right of the plane wave operator. We have multiplied and divided by $i E$ so that sigmoid is replaced with its time derivative. Plugging this in to \eq{I1} we can relate to the function  $F(k^1,...,k^5)$:
\begin{equation}
I_1 = \int \frac{dE}{2\pi}\dot{\sigma}(E)\int_0^1 \frac{dy}{y}2\pi\delta(E) F(-i,-i,E,i,i)\left(\frac{2\sinh\big((2i+\frac{1}{2}E)E\ln y\big)}{iE}\right).
\end{equation}
There is potential ambiguity in multiplication of distributions. It is important that the hyperbolic sine is canceled with the pole at $E=0$ and not multiplied with the delta function. Using
\begin{equation}
\lim_{E\to 0} \frac{2\sinh\big((2i+\frac{1}{2}E)E\ln y\big)}{iE} = 4\ln y,
\end{equation}
we can integrate the delta function to obtain 
\begin{equation}
I_1 = 4 \int_0^1 \frac{dy}{y}\ln y \, F(-i,-i,0,i,i).
\end{equation}
The sigmoid drops out because $\dot{\sigma}(0)=1$. In a similar way we can compute the expressions
\begin{equation}
I_2 = \frac{1}{V}\left\langle  \sigma_+(\Psi_+,\Psi_+),\frac{b_0}{L_0}\Psi_-^2\right\rangle,\ \ \ \ I_3= \frac{1}{V}\left\langle \sigma_+(\Psi_-,\Psi_-),\frac{b_0}{L_0}\Psi_+^2\right\rangle,
\end{equation}
giving
\begin{equation}
I_2 = -I_3 = 2 \int_0^1 \frac{dy}{y}\left.\left(4\ln y +i\frac{d}{ds}\right) F(-i+s,-i+s,0,i,i)\right|_{s=0}
\end{equation}
as the result.

The final term in $\Omega_3$ is
\begin{equation}
I_4 = \frac{1}{V}\left\langle \sigma_-(\Psi_+,\Psi_-),\frac{b_0}{L_0}[\Psi_+,\Psi_-]\right\rangle.
\end{equation}
Here we need to be careful with the definition of the propagator because it acts on a state which does not necessarily have positive conformal weight. This is because the momentum cancels between $\Psi_+$ and $\Psi_-$, rather than adding up as with previous terms. The Schwinger representation may encounter divergences from tachyonic states, where the mode counting operator $N$ vanishes, and from massless states, where the mode counting operator is one. However, massless states will not be present because the rolling tachyon solution is exactly marginal. Therefore we only need to worry about the tachyon divergence. To deal with this we add and subtract a projector onto the tachyon state: 
\begin{equation}
I_4 = \frac{1}{V}\left\langle \sigma_-(\Psi_+,\Psi_-),\frac{b_0}{L_0}\delta_N [\Psi_+,\Psi_-]\right\rangle+\frac{1}{V}\left\langle \sigma_-(\Psi_+,\Psi_-),\frac{b_0}{L_0}(1-\delta_N) [\Psi_+,\Psi_-]\right\rangle.
\end{equation}
In the first term we know that $L_0=-1$ because $N=0$ and there is no momentum flowing through the propagator. In the second term we can define the propagator through the Schwinger representation because the tachyon state is projected out. In this way we find
\begin{align}
I_4=-\frac{1}{V}\big\langle \sigma_-(\Psi_+,\Psi_-),b_0\delta_N [\Psi_+,\Psi_-]\big\rangle+\frac{1}{V}\int_0^1 \frac{dy}{y}\bigg[& \big\langle \sigma_-(\Psi_+,\Psi_-),b_0 y^{L_0}[\Psi_+,\Psi_-]\big\rangle\nonumber\\ & -\frac{1}{y}\big\langle \sigma_-(\Psi_+,\Psi_-),b_0 \delta_N [\Psi_+,\Psi_-]\big\rangle\bigg].
\end{align}
The tachyon contribution can be computed from the overlap
\begin{align}
&\Big\langle c_1|k^1\rangle*c_1|k^2\rangle,\,  b_0\delta_N \Big(c_1|k^4\rangle*c_1|k^5\rangle\Big)\Big\rangle \nonumber\\
&\ \ =\kappa^2(2\pi)^{26}V\delta^{26}(k^{12}+k^{45})\exp\left(-\frac{1}{2}N_{00}^{11}\Big((k^1)^2+(k^2)^2+(k^4)^2+(k^5)^2+2(k^{12})^2\Big)\right),
\end{align}
where $N_{00}^{11}=\ln(27/16)$. We find
\begin{equation}
\frac{1}{V}\big\langle \sigma_-(\Psi_+,\Psi_-),b_0\delta_N [\Psi_+,\Psi_-]\big\rangle = -8\kappa^2 N_{00}^{11} e^{-2N_{00}^{11}}.
\end{equation}
Therefore 
\begin{align}
I_4 = 8\kappa^2 N_{00}^{11} e^{-2N_{00}^{11}}+ & \int_0^1\frac{dy}{y} \Bigg[\frac{8\kappa^2 N_{00}^{11} e^{-2N_{00}^{11}}}{y}\nonumber\\
& - \sum_{\alpha,\beta=\pm 1}i\alpha\frac{d}{ds}\Big(F(-i\alpha+s,i\alpha,0,-i\beta,i\beta)-F(-i\alpha,i\alpha+s,0,-i\beta,i\beta)\Big)\Bigg],
\end{align}
where the remaining term in integrand has been expressed in terms of $F(k^1,...,k^5)$.

\subsection{Results}

Here we present the numerical results for the terms $I_1, I_2=-I_3$, and $I_4$ following the oscillator truncation method of Taylor. The details of our implementation can be found in the \texttt{Mathematica} document accompanying this paper.

\begin{figure}[t]
	\centering
	\includegraphics[scale=.6]{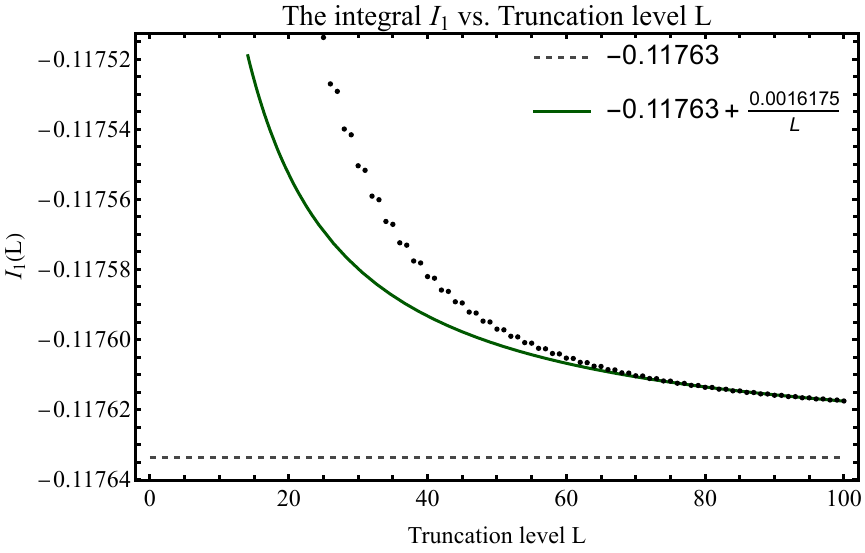}
	\includegraphics[scale=.575]{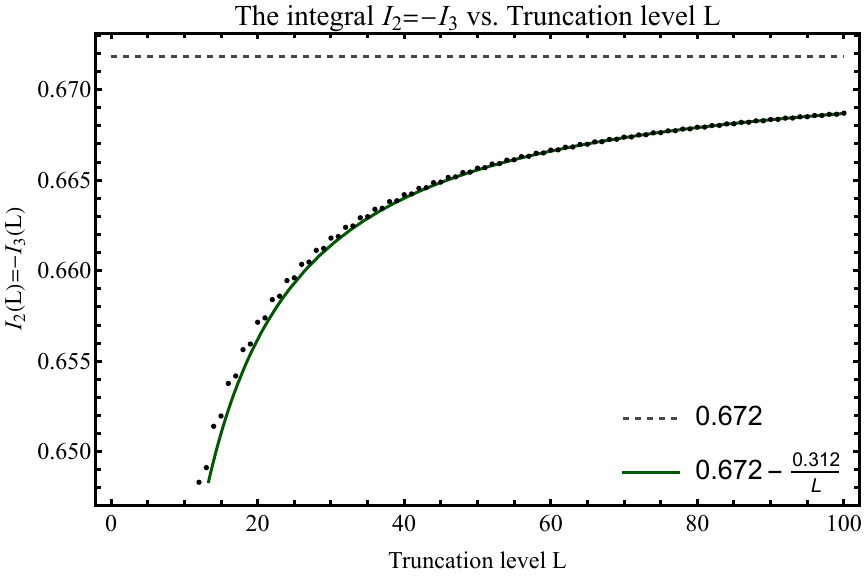}
	\includegraphics[scale=.6]{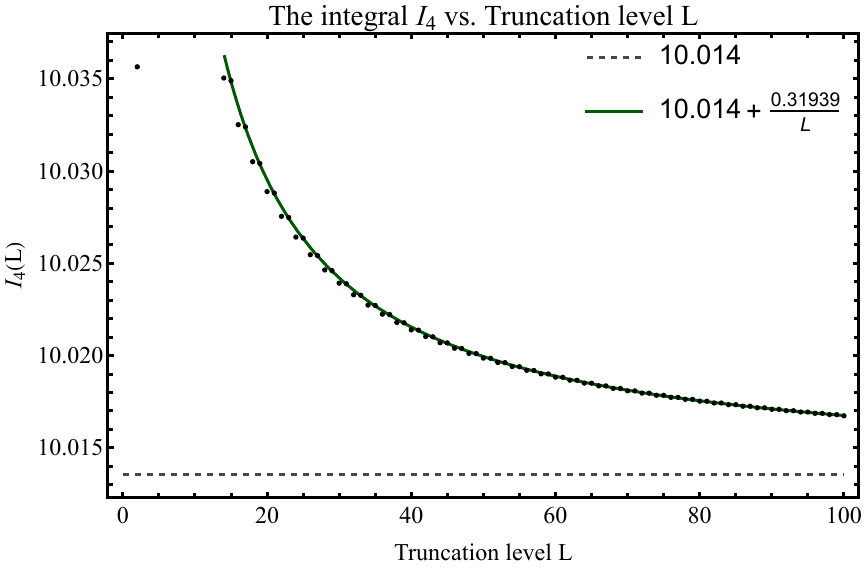}
	\caption{The oscillator level-truncated approximations to the contributions $I_1, I_2=-I_3$, and $I_4$ to the symplectic form at third order, $\Omega_3$.}\label{fig:SympOSFT4}
\end{figure} 

The prescription is to place an upper limit $L$ on the mode number index, so that the Neumann coefficients and their modified versions \eq{modified_NC} are truncated to finite $L\times L$ matrices. All products, determinants, and inverses in \eq{4tach_matrix} and \eq{QAB} are evaluated numerically with the truncated matrices. The resulting approximation to $I_1,...,I_4$ is shown in figure~\ref{fig:SympOSFT4} up to $L=100$. As observed by Taylor \cite{Taylor}, the truncation introduces errors of order $1/L$. Performing a least-square fit for the last $25$ terms, the approximation to $I_1,...,I_4$ may be presented as an expansion in powers of $1/L$ 
\begin{subequations}
	\begin{align}
		&I_1(L) = -0.117634 + {0.00161755 \over L } + \cdots, \\
		&I_2(L) = 0.671804 - { 0.311937 \over L } + \cdots, \\
		&I_4(L) = 10.0136 + { 0.319393 \over L} + \cdots.
	\end{align}
\end{subequations}
The leading term is the predicted exact result for $L\to\infty$. The third order coefficient of the symplectic structure is given as 
\begin{align} \label{eq:Omega3}
	\frac{\Omega_3(L)}{V/g^2} & =
	{1 \over 2} I_1(L) - {1 \over 8} I_2(L) +  {1 \over 8} I_3(L) +  {1 \over 4} I_4(L) \nonumber\\
	&= 2.27662 + {0.158641 \over L} + \cdots.
\end{align}
Extracting the $L\to\infty$ prediction and dividing by four, we find 
\begin{equation}
E(\lambda) \approx \frac{V}{g^2}\left(-\frac{1}{2}\lambda^2 + 0.569155\, \lambda^4+ \mathcal{O}(\lambda^6)\right)
\end{equation}
as the energy of the rolling tachyon solution.

The energy can alternatively be computed from the boundary state, which can be obtained from the solution via the Ellwood invariant \cite{Ellwood2,Kudrna2}. The energy is
\begin{equation}E(\lambda) = \frac{i}{\pi g^2 V_{X^0}}W(D,\Psi),\end{equation}
where 
\begin{equation}
W(D,\Psi) = \big\langle I, D(i,\overline{i})\Psi\big\rangle
\end{equation}
is the Ellwood invariant computed from the matter dilaton state,
\begin{equation}D(z,\overline{z}) = c \overline{c} \partial X^0 \overline{\partial}X^0(z,\overline{z}),\end{equation}
and $I$ is the identity string field. Kudrna evaluated a related quantity for the cosine deformation in \cite{Kudrna} (see his observable $D_1$ in table 6.3), whose inverse Wick rotation gives the energy of the rolling tachyon solution as
\begin{equation}
E(\lambda) \approx \frac{V}{g^2}\left(-\frac{1}{2}\lambda^2 + 0.5689\, \lambda^4+ \mathcal{O}(\lambda^6)\right),\ \ \ \ \text{(Kudrna \cite{Kudrna})}.
\end{equation}
Our coefficient at order $\lambda^4$ agrees within an accuracy of $0.04$ percent. Cho, Mazel, and Yin computed the energy from their version of the symplectic structure, and also independently from the Ellwood invariant. They reported (see their equations (4.17) and (4.21))
\begin{subequations}
\begin{align}
E(\lambda) &\approx \frac{V}{g^2}\left(-\frac{1}{2}\lambda^2 + 0.5630\, \lambda^4+ \mathcal{O}(\lambda^6)\right),\ \ \ \ \ \ \ \begin{matrix} \text{(Cho, Mazel, Yin \cite{Cho}} \\ \text{from symplectic form)}\end{matrix},\\
E(\lambda) & \approx \frac{V}{g^2}\left(-\frac{1}{2}\lambda^2 + 0.5595\, \lambda^4+ \mathcal{O}(\lambda^6)\right),\ \ \ \ \ \ \ \begin{matrix} \text{(Cho, Mazel, Yin \cite{Cho}} \\ \text{from Ellwood invariant)}\end{matrix}.
\end{align}
\end{subequations}
Our results more closely match Kudrna's, but they all show reasonable agreement within roughly 2 percent. 

Cho, Mazel and Yin have demonstrated analytically that the energy computed from the Ellwood invariant must agree to all orders in $\lambda$ with that computed from their version of the symplectic structure. The formal equivalence between their symplectic structure and ours is explained in more detail below. The numerical results demonstrate that these arguments can be trusted. 

\subsection{Relation to Cho, Mazel, and Yin}
\label{subsec:Cho}

It has often been observed that Witten's SFT is formally local if the string field is expressed as a function of the position of the string midpoint \cite{Witten2}. Therefore, in principle we can find the symplectic structure in the same way as we would in any local field theory, writing coordinates on an $X^0(i,\overline{i})=0$ time slice and computing their canonical momenta from the Lagrangian. The result is the symplectic structure proposed by Witten \cite{Witten}, who wrote it as 
\begin{equation}\Omega = -\frac{1}{2g^2}\Big\langle \delta\Psi,\big[Q,\Theta\big(X^0(i,\overline{i})\big)\big]\delta\Psi\Big\rangle,\end{equation}
where 
\begin{equation}
\Theta\big(X^0(i,\overline{i})\big) =\int \frac{dE}{2\pi}\frac{1}{iE} e^{i E X^0(i,\overline{i})}.\label{eq:thetasigma}
\end{equation}
The expression clearly fits our general formula \eq{Omega}. As mentioned earlier, the interacting contribution drops out because the midpoint coordinate commutes through the star product. Unfortunately, Witten's symplectic structure is undefined. The problem comes from the fact that the plane wave vertex operators in the sigmoid have nonzero conformal weight. They obey the transformation~law 
\begin{equation}
f\circ e ^{i E X^0(i,\overline{i})} = |\d f(i)|^{-\frac{E^2}{2}} e ^{i E X^0(f(i),\overline{f(i)})}.
\end{equation}
The conformal maps relevant to Witten's SFT usually have conical singularities at the midpoint. This means that $\d f(i)$ will be vanishing or infinite, making the symplectic structure undefined. The difficulty can be traced back to the assumption that Witten's SFT is local in the midpoint coordinate. Attempts to make this locality manifest encounter serious difficulties from short distance singularities on the worldsheet \cite{Morris,Manes,Potting,Belov}.

To avoid these issues, Cho, Mazel, and Yin propose to define the symplectic structure by projecting Witten's sigmoid to its weight zero part. The result is 
\begin{equation}
\Omega = -\frac{1}{2g^2 V_{X^0}}\big\langle \delta\Psi,\gamma^0(i,\overline{i})\delta\Psi\big\rangle,\label{eq:CMY}
\end{equation}
where $V_{X^0}$ is the (infinite) volume of the time coordinate and
\begin{equation}\gamma^\mu(z,\overline{z}) = c\d X^\mu(z) + \overline{c}\overline{\d}X^\mu(\overline{z})\end{equation}
is a bulk insertion of the photon vertex operator. Interestingly, the volume divergence is not regularized and there is no time slice, even in the looser sense of the sigmoid. Nevertheless this formula correctly reproduces the rolling tachyon energy. 

We would like to understand how Cho, Mazel and Yin's formula relates to our approach. For this we return to Witten's symplectic structure. The difficulty with the original proposal is that the timelike free boson has nontrivial self-contractions. To avoid this we work with a lightlike free boson
\begin{equation}X^+(z,\overline{z}) = \frac{1}{\sqrt{2}}\Big(X^0(z,\overline{z})+ X^1(z,\overline{z})\Big),\label{eq:Xplus}\end{equation}
where we assume $X^1(z,\overline{z})$ is spacelike and can be found in the spatial BCFT. We consider a sigmoid 
\begin{equation}
\sigma = \int \frac{dE}{2\pi} \sigma(E) e^{iE  X^+(i,\overline{i})}.\label{eq:mplc_sigmoid}
\end{equation}
Because the lightlike plane wave vertex operator has vanishing conformal weight, conformal transformation does not lead to singularity. The symplectic structure appears to be well-defined and takes the form 
\begin{equation}
\Omega = -\frac{1}{2g^2}\big\langle \delta\Psi,\gamma^+(i,\overline{i})\dot{\sigma}\,\delta\Psi\big\rangle,\label{eq:mplc_Omega}
\end{equation}
where the dot on $\sigma$ indicates the derivative with respect to $X^+(i,\overline{i})$. The success of this construction is related to the fact that Witten's SFT is local in a more genuine sense when considering only the lightlike coordinate of the string midpoint. This is the basis of the canonical formulation of Witten's SFT in midpoint lightcone time \cite{Maeno,Erler2}. In that context $X^1(z,\overline{z})$ is a noncompact free boson subject to Neumann boundary conditions, but we consider Dirichlet boundary conditions so that we do not have to worry about the zero mode. It is interesting to mention that the midpoint-lightcone sigmoid does not act through multiplication by a function $\sigma(x)$. It mixes fields of different mass level and operates on Lorentz indices. This shows how the sigmoid can usefully generalize the concept of time slice. 

We can arrive at Cho, Mazel, and Yin's symplectic structure as follows. Write \eq{mplc_sigmoid} as a correlation function on the upper half plane:
\begin{equation}
\Omega = -\frac{1}{2 g^2}\int\frac{dE}{2\pi}\dot{\sigma}(E)\Big\langle\big(I\circ\delta\Psi(0)\big)\delta\Psi(0)\, \gamma^+e^{iE X^+(i,\overline{i})}\Big\rangle_\UHP,
\end{equation}
where $I(z)=-1/z$ is the BPZ conformal map. Because the symplectic structure is supposed to be independent of the sigmoid, we know that only insertions at zero energy contribute to the correlation function. Therefore 
\begin{equation}
\Omega = -\frac{1}{2 g^2}\int\frac{dE}{2\pi}\dot{\sigma}(E)\Big\langle\Big[\big(I\circ\delta\Psi(0)\big)\delta\Psi(0)\gamma^+(i,\overline{i})\Big]_{E=0}e^{iE X^+(i,\overline{i})}\Big\rangle_\UHP,\label{eq:ins_0m}
\end{equation}
where $[\cdots]_{E=0}$ indicates a projection to the zero energy part of the insertions. If the insertions have complicated or singular behavior near zero energy there might be some doubt as to how this projection should be defined. We will not worry about this because for rolling tachyon solutions the meaning is clear. Next we observe that for any insertions $\mathcal{O}$ at zero energy we have the identity
\begin{equation}
\big\langle \mathcal{O}\, e^{i E X^+(z,\overline{z})}\big\rangle_\UHP = 
\big\langle \mathcal{O}\big\rangle_\UHP\frac{1}{\big\langle 1\big\rangle_\UHP^\mathrm{matter}}\big\langle e^{i E X^+(z,\overline{z})}\big\rangle_\UHP^\mathrm{matter}.\label{eq:factorize}
\end{equation}
Because $\mathcal{O}$ has zero energy the correlation function is nonzero only when the plane wave vertex operator reduces to the identity. Then all contractions between $\mathcal{O}$ and the plane wave vertex operator vanish, and the correlation function effectively factorizes. Since we assume that the spatial component of $X^+$ is subject to Dirichlet boundary conditions we have
\begin{equation}
\big\langle e^{i E X^+(z,\overline{z})}\big\rangle_\UHP^\mathrm{matter} = 2\pi V\delta\left(\frac{E}{\sqrt{2}}\right),
\end{equation}
where the delta function comes from the time direction which is subject to Neumann boundary conditions and $\sqrt{2}$ comes from \eq{Xplus}. Also using
\begin{equation}
\big\langle 1\big\rangle_\UHP^\mathrm{matter} = V V_{X^0},
\end{equation}
we can plug in to \eq{ins_0m} to find 
\begin{align}
\Omega = -\frac{1}{2g^2 V_{X^0}}\int \frac{dE}{2\pi}\dot{\sigma}(E)2\pi \delta\left(\frac{E}{\sqrt{2}}\right)
\Big\langle\Big[\big(I\circ\delta\Psi(0)\big)\delta\Psi(0)\gamma^+(i,\overline{i}) \Big]_{E=0} \Big\rangle_\UHP.
\end{align}
Now it is easy to integrate over the energy using $\dot{\sigma}(0)=1$ to obtain
\begin{align}
\Omega = -\frac{1}{\sqrt{2}g^2 V_{X^0}}
\Big\langle\Big[\big(I\circ\delta\Psi(0)\big)\delta\Psi(0)\gamma^+(i,\overline{i}) \Big]_{E=0} \Big\rangle_\UHP.
\end{align}
The sigmoid has disappeared. Now let us assume solutions that do not excite primaries in the $X^1$ BCFT. Then the spatial component of $\gamma^+(i,\overline{i})$ drops out leaving 
\begin{align}
\Omega = -\frac{1}{2 g^2 V_{X^0}}
\Big\langle\Big[\big(I\circ\delta\Psi(0)\big)\delta\Psi(0)\gamma^0(i,\overline{i})\Big]_{E=0} \Big\rangle_\UHP.
\end{align}
The final step is to observe that the projection onto zero energy is no longer necessary because it is enforced automatically by momentum conservation. Then we recover \eq{CMY}. The conclusion is that Cho, Mazel, and Yin's formula can be seen as a reduction of the midpoint-lightcone symplectic form. Independence from the choice of sigmoid then explains why our results must agree with theirs. Our argument assumes that the worldsheet theory has a spectator free boson $X^1$ whose primaries are not excited by the solution. This will not be true, for example, on a ZZ-brane in $c=1$ string theory. It may be possible to strengthen the derivation by considering Witten's original sigmoid as a limit. 

We can ask how our computation of the rolling tachyon energy compares to Cho, Mazel and Yin's. Their approach has the advantage that interacting contributions to the symplectic structure drop out. However, there are additional contributions from $\Psi_1$-$\Psi_3$ terms which do not vanish because the sigmoid does not commute through the mode counting operator. From a numerical point of view both approaches seem to be viable, though our different methods of evaluation make a precise comparison difficult. The conclusion is that our calculation of the symplectic structure is quite simple and natural. Perhaps this is a surprise given the seeming awkwardness of the position zero mode from the point of view of the geometry of the Witten~vertex. Therefore we do not expect important obstacles to performing similar calculations to other string field theories, most importantly closed~SFT. 

The main advantage of the midpoint-lightcone formulation is that it is connected to an initial value problem. If the initial value problem is viable, it should give a systematic parametrization of the phase space of open string field theory. Whether it really does this remains to be seen. Another path to constructing the phase space is though lightcone gauge, but this approach is complicated by the appearance of singular Mandelstam diagrams in the process of gauge fixing~\cite{Erler5,Erler8}. The explicit construction of phase space in SFT remains an important open problem. 

\subsection*{Acknowledgments}

TE and AHF would like to thank D. Gross for conversations and hospitality at the KITP while carrying out part of this work. AHF would like to thank M. Rangamani and B. Zwiebach for discussions. We also thank A. Sen for comments on the draft. The work of AHF is supported by the U.S. Department of Energy, Office of Science, Office of High Energy Physics of U.S. Department of Energy under grant Contract Number DE-SC0009999, and the funds from the University of California. The work of VB and TE was supported by the European Structural and Investment Funds and the Czech Ministry of Education, Youth and Sports (project No. FORTE—CZ.02.01.01/00/22\_008/0004632). This research was supported in part by grant NSF PHY-2309135 and the Gordon and Betty Moore Foundation Grant No. 2919.02 to the Kavli Institute for Theoretical Physics~(KITP).

\end{document}